\let\KernelLabel\label
\documentclass[prb,twocolumn,aps,superscriptaddress,10pt,showkeys]{revtex4-2}
\let\label\KernelLabel
\usepackage{graphicx}
\usepackage{array}[=2016-10-06]
\usepackage{dcolumn}
\usepackage{amsmath}
\usepackage{amssymb}
\usepackage{bm}
\usepackage{mathtools}
\usepackage{physics}
\usepackage[utf8]{inputenc}
\usepackage[dvipsnames]{xcolor}
\usepackage[english]{babel}
\usepackage[T1]{fontenc}
\usepackage{microtype}
\usepackage{bbm}
\usepackage{verbatim}
\usepackage[caption=false]{subfig}
\usepackage{xfrac}
\usepackage{upgreek}
\usepackage{makecell}
\usepackage{comment}
\usepackage{placeins}
\usepackage[normalem]{ulem}
\usepackage{hyperref}
\hypersetup{colorlinks = true, linkcolor=blue, citecolor=blue, urlcolor=blue}

  \makeatletter
    \renewcommand\@make@capt@title[2]{%
     \@ifx@empty\float@link{\@firstofone}{\expandafter\href\expandafter{\float@link}}%
      {\textsc{#1}}\@caption@fignum@sep#2\quad}%
    \makeatother

\newcommand{\panel}[3]{%
  \begin{minipage}[t]{#1}%
    \raggedright\textbf{\footnotesize (#2)}\\[-2pt]%
    \includegraphics[width=\linewidth]{#3}%
  \end{minipage}%
}

\begin{document}

\title{Sub-Thermionic Switching in a Negative-Capacitance Bandgap-Change Field-Effect Transistor}

\author{Rahul B. Awale}
\affiliation{Department of Electrical Engineering, Indian Institute of Technology Bombay, Powai, Mumbai-400076, India}
\affiliation{Department of Materials Science and Engineering, Monash University, Clayton, Victoria 3800, Australia}

\author{Mike Klymenko}
\affiliation{Department of Materials Science and Engineering, Monash University, Clayton, Victoria 3800, Australia}

\author{Yuefeng~Yin}
\affiliation{Department of Materials Science and Engineering, Monash University, Clayton, Victoria 3800, Australia}

\author{Nikhil~V~Medhekar}
\affiliation{Department of Materials Science and Engineering, Monash University, Clayton, Victoria 3800, Australia}
\affiliation{School of Physics and Astronomy, Monash University, Clayton, Victoria 3800, Australia}

\author{Bhaskaran~Muralidharan}
\affiliation{Department of Electrical Engineering, Indian Institute of Technology Bombay, Powai, Mumbai-400076, India}

\author{Michael~S.~Fuhrer}
\email{Corresponding author: michael.fuhrer@monash.edu}
\affiliation{School of Physics and Astronomy, Monash University, Clayton, Victoria 3800, Australia}
\affiliation{Department of Materials Science and Engineering, Monash University, Clayton, Victoria 3800, Australia}

\keywords{bilayer graphene, negative capacitance bandgap change field effect transistor, ferroelectric, sub-thermionic switching, subthreshold swing, AlScN}
\date{\today}

\begin{abstract}
Overcoming the approximately 60~mV/dec Boltzmann limit of the subthreshold swing remains a central challenge for low-power nanoelectronics. Negative-capacitance field-effect transistors address it by amplifying the channel potential, but their effectiveness is constrained by capacitance matching and by the quantum capacitance of conventional channels. Here we model a negative-capacitance bandgap-change field-effect transistor (NC-BCFET), in which the ferroelectric gate stack instead amplifies the electric field that opens the bandgap of the channel, using experimentally proven materials: bilayer graphene as a channel with an electrically tunable bandgap and Al$_{0.55}$Sc$_{0.45}$N as the negative-capacitance ferroelectric. Our self-consistent framework couples a four-band tight-binding Hamiltonian with GW-corrected screening, a quasistatic Landau--Devonshire ferroelectric response, and ballistic Landauer--Büttiker transport. Negative capacitance amplifies the interlayer potential difference, compressing the switching window 37-fold relative to a dielectric-gated control. At 300~K, the optimized NC-BCFET achieves a subthreshold swing of 15~mV/dec, four times below the Boltzmann limit, improving to 2.5~mV/dec at 100~K. The room-temperature on/off ratio is limited to $\sim 10^{2}$ by the bandgap of bilayer graphene. The concept extends directly to materials with larger field-tunable bandgaps, where deep subthermionic switching and high on/off ratios can be achieved together at room temperature.
\end{abstract}
\maketitle

\section{\label{sec:intro}Introduction}

Power dissipation has become the dominant constraint on the continued scaling of CMOS technologies~\cite{Gargini2020IRDS}. The IEEE International Roadmap for Devices and Systems (IRDS) identifies steep-slope devices that switch with less gate voltage than a conventional transistor as a central objective for low-power electronics beyond CMOS~\cite{Gargini2020IRDS}. Reducing the supply voltage lowers the energy per switching operation, but only if the transistor can still turn off cleanly over a small gate-voltage range. That requirement is set by the {\it subthreshold swing}, and its Boltzmann limit of approximately 60 mV/dec of current at room temperature is the barrier that steep-slope devices must overcome~\cite{Rahman2003,Beckers2020}.

\indent The subthreshold swing (SS)~\cite{Rahman2003} is defined as
\begin{equation}
SS = \left[\frac{\partial \log_{10}(I_{\mathrm{DS}})}{\partial V_g}\right]^{-1}.
\label{Eq:ssdef}
\end{equation}
In a conventional MOSFET with standard dielectric materials, we expect
\begin{equation}
SS = \text{ln}(10)\,(k_BT/q)\,(1 + C_s/C_{\mathrm{ins}}).
\label{Eq:eq1}
\end{equation}
Here $m=dV_g/d\psi_s=1+C_s/C_{\mathrm{ins}}$ is the body factor and $n=d\psi_s/d\log_{10}(I_{\mathrm{DS}})$ is the transport factor, so the general identity is $SS=mn$. For ideal thermionic injection, $n\approx\ln(10)k_BT/q$, which gives Eq.~(\ref{Eq:eq1}) and limits a conventional room-temperature MOSFET to approximately 60~mV/decade~\cite{Beckers2020}.\\
\indent Several approaches have been proposed to overcome this limit, including tunnel FETs~\cite{Cao2020,Cao2014,Ionescu2011} and negative-capacitance FETs~\cite{Salahuddin2008,Wong2019} in which a ferroelectric layer in the gate stack provides voltage amplification and reduces the body factor $m$ below unity. Experimental NC-FET studies have reported sub-60~mV/dec operation~\cite{McGuire2017,Khan2015}; however, Cao and Banerjee~\cite{Cao2020} emphasize that the rapidly increasing quantum capacitance near threshold restricts the positive-total-capacitance design window and complicates hysteresis-free operation, which makes robust subthermionic switching impractical in conventional NC-FET designs.
\\
\begin{figure*}[tbp]
    \centering
    \panel{0.4950\textwidth}{a}{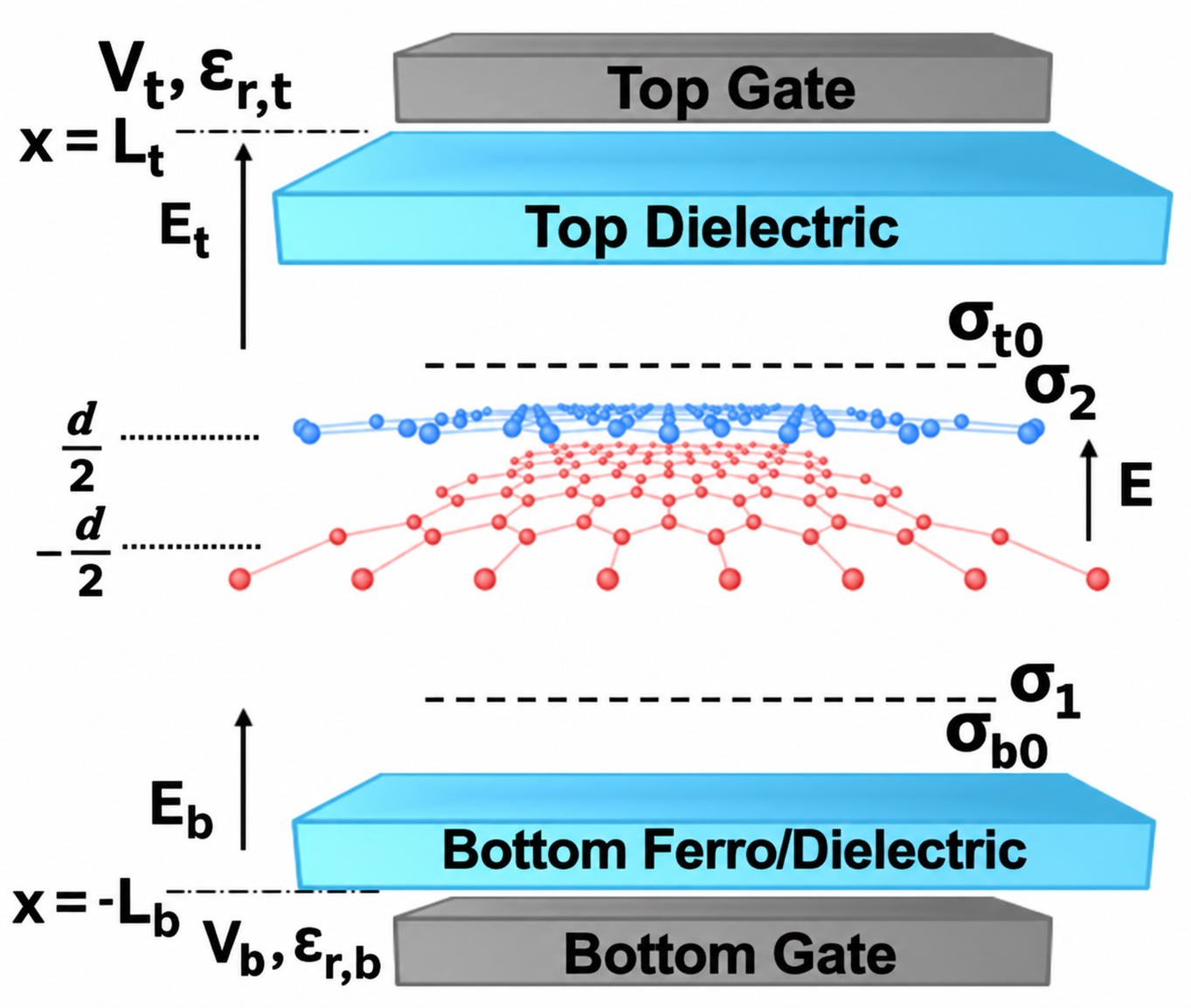}
    \hfill
    \panel{0.4950\textwidth}{b}{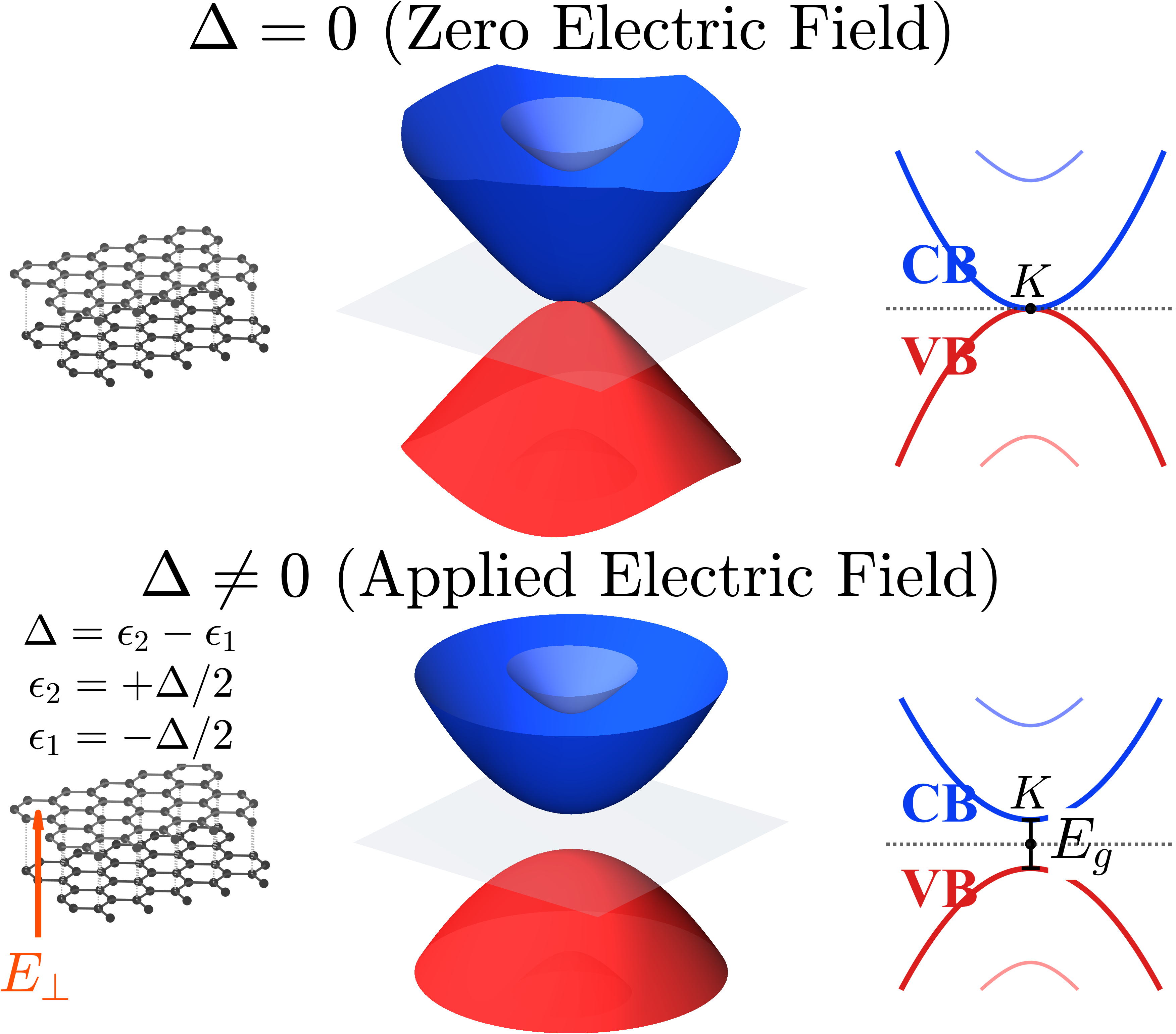}

    \caption{(a) Electrostatics of the gated BLG lattice structure showing the upper (blue) and lower (red) graphene layers with charge densities $\sigma_{2}=-en_{2}$ and $\sigma_{1}=-en_{1}$, respectively. Here, BLG is intrinsic; the equilibrium reference charge densities vanish, such that $\sigma_{t0}=-en_{t0}=0$ and $\sigma_{b0}=-en_{b0}=0$. $E_t$ and $E_b$ are the electric fields inside the top and bottom dielectrics, respectively, while $d$ is the interlayer spacing of the BLG channel. And $\varepsilon_{r,t}$ and $\varepsilon_{r,b}$ are the permittivities of the top (with thickness $L_t$) and bottom (with thickness $L_b$) dielectrics, respectively. (b) Low-energy band structure of BLG. Without a perpendicular electric field $E_{\perp}$ ($\Delta=0$), the conduction and valence bands touch at the $K$ point; an applied field creates the interlayer onsite-energy asymmetry $\Delta=\epsilon_{2}-\epsilon_{1}$~\cite{McCann2013,McCann2006, McCann2007} and opens a tunable band gap $E_g$. Here, $\epsilon_{2}$ and $\epsilon_{1}$ are the onsite energies of graphene layers 2 and 1, respectively.}
    \label{fig:1}
\end{figure*}
\indent An alternative proposal, the negative capacitance topological quantum FET (NC-TQFET)~\cite{fuhrer2021nctqfet}, couples a ferroelectric NC gate to a channel material in which the bandgap can be varied by an electric field~\cite{Nadeem2021,Nadeem2022}. While Ref.~\citenum{fuhrer2021nctqfet} focused on materials near a topological phase transition, here we generalize this concept to any bandgap-change (BC) channel material to consider the negative-capacitance bandgap-change FET (NC-BCFET). In the NC-BCFET, the NC stack directly drives bandgap modulation by \textit{amplifying the electric field in the channel}, rather than merely lowering the body factor, so that the gate-tunable bandgap, not the channel capacitance alone, sets the switching. 
Although Ref.~\citenum{fuhrer2021nctqfet} set out the basic principles of the NC-BCFET concept, including bilayer graphene (BLG) as a possible BC channel, a detailed device model that encompasses realistic materials and treats the electrostatics of the NC gate stack and channel is lacking.
\\
\indent Here, we develop a self-consistent device model for a BLG NC-BCFET, in which the experimentally proven BC material BLG serves as the channel~\cite{McCann2013,McCann2006,CastroNeto2009,Zhi2020,Castro2007}. The dual-gate architecture retains a top dielectric while replacing the bottom dielectric with a ferroelectric Al$_{0.55}$Sc$_{0.45}$N layer. The electronic structure is described by a GW-corrected tight-binding model~\cite{Gava2009}, the ferroelectric by the quasistatic equilibrium limit of a sixth-order Landau--Devonshire (L-D) model~\cite{Salahuddin2008,Wong2019,Khan2015}, and finite-bias transport by the Landauer--B\"uttiker (L-B) framework~\cite{Rahman2003,Lundstrom2017}. The quantum capacitance is retained in full: it is not imposed as a lumped element but enters through the density-of-states-weighted charge integral of the self-consistent channel potential (Methods).
\\
\indent Relative to a BLG BCFET control, the calculated BLG NC-BCFET exhibits a compressed switching window and room-temperature subthermionic switching, with further sharpening at 100~K. From the direct simulated drain currents, the lowest subthreshold swing $SS_{\min}$ is 15~mV/dec at 300~K for the optimal ferroelectric thickness $L_b=10$~nm, and 2.5 mV/dec at 100 K for $L_b=20$~nm; the BLG BCFET (oxide device) remains above 900~mV/dec over the examined drain-bias range. 
Our results show the robustness of the BLG NC-BCFET concept within a detailed device model utilizing realistic materials. The findings should be generally applicable to other BC channel materials with large bandgaps, suitable for steep-slope logic devices with high on-off ratios at room temperature.

\section{\label{sec:results}Results and discussion}

\indent The BLG NC-BCFET analyzed in this work contains a BLG channel between a top dielectric and a bottom ferroelectric layer, which separate the channel from the metal top and bottom gates. For benchmarking, we also simulate a reference BLG BCFET in which the bottom ferroelectric is replaced by dielectric SiO$_2$. We choose HfO$_{2}$ for the dielectric, and Al$_{0.55}$Sc$_{0.45}$N for the ferroelectric for its CMOS compatibility, large remnant polarization, III--V process integration, and well-characterized L-D parameters~\cite{Gu2024}. 
Figure~\ref{fig:1}(a) shows the one-dimensional gate-stack electrostatics. The BLG channel is grounded, the top-gate voltage $V_{t}$ is fixed, and the bottom-gate voltage $V_{b}$ is varied. Layer charges follow from Gauss's law and displacement-field continuity. The ferroelectric is represented by a quasistatic sixth-order Landau free energy whose negative-curvature interval gives a negative differential capacitance~\cite{Salahuddin2008,Wong2019}. A perpendicular field creates the interlayer asymmetry $\Delta$, lifts the K-point degeneracy, and opens the BLG transport gap~\cite{Ohta2006} [Figure~\ref{fig:1}(b)]. Intralayer screening absent from a non-interacting TB treatment is incorporated with the density-dependent \textit{ab initio} GW parametrization of Gava \textit{et al.}~\cite{Gava2009}. The coupled polarization, layer charges, GW screening, bandgap, and channel potential are then solved self-consistently (Supporting Information, Sections~1--3).


\indent Optical~\cite{Mak2009,Zhang2009} and electronic~\cite{Yan2010} measurements have shown that the electronic charge distribution in BLG under an applied electric field partially cancels the field, reducing the effective $\Delta$ and the bandgap. Gava \textit{et al.}~\cite{Gava2009} identified two types of screening: interlayer and intralayer. Interlayer screening occurs under a perpendicular external field where the two layers are held at different potentials ($\sigma_2 - \sigma_1 \neq 0$), compensating the applied field, which is captured in the standard TB model. Intralayer screening arises from charge redistribution within each graphene layer. 
Using density functional theory (DFT), Gava \textit{et al.}~\cite{Gava2009} showed that the linearly induced charge density $\rho^{(1)}(z)$ in each layer can be decomposed into a symmetric component $\rho_s$, related to interlayer screening, and an antisymmetric component $\rho_a$, related to intralayer screening. The antisymmetric component is absent in the TB model, which treats each layer as an infinitely thin sheet incapable of internal polarization. We therefore employ the DFT-derived parameters of Gava \textit{et al.}~\cite{Gava2009} to capture the full impact of screening on the bandgap.

\indent Charge transport is treated within the ballistic top-of-the-barrier (ToB) L-B model~\cite{Rahman2003,Mayorov2011}, with the source and drain held at separate electrochemical potentials; in contrast to a conventional MOSFET, the bandgap itself depends strongly on $V_b$ and becomes important in determining switching. A central quantity in this description is the channel band shift, $\Delta E = E_F - E_{\mathrm{CNP}}$. The source contact is the grounded reference, so its electrochemical potential is pinned and it is the BLG bands that move: gate-induced charging of the bilayer displaces the charge-neutrality point rigidly with respect to $E_F$~\cite{Zhi2020}. We do not subtract $\Delta E$ from the gate voltage drops that set the layer charge densities; the entire common mode is instead carried by the ToB channel potential $U$, which appears only inside the Fermi functions, so that $\Delta E = -U$.
The ferroelectric polarization $P_f$, interlayer asymmetry $\Delta$, carrier density $n$, and channel potential $U$ are obtained from a single self-consistent calculation.
Because $U$ shifts the channel bands relative to the pinned reservoir level $E_F = \mu_S = 0$, and so changes $\Delta E$, it also changes the carrier density and, through it, the GW screening coefficient $\alpha_{\mathrm{GW}}(n)$, the interlayer asymmetry $\Delta$, and the ferroelectric polarization $P_f$. The transport-level electrostatics feed back into the band structure at every bias point. Full details are given in the Methods section and in Section~1.1 of the Supporting Information.

\begin{figure*}[tbp]
    \centering
    \panel{0.3885\textwidth}{a}{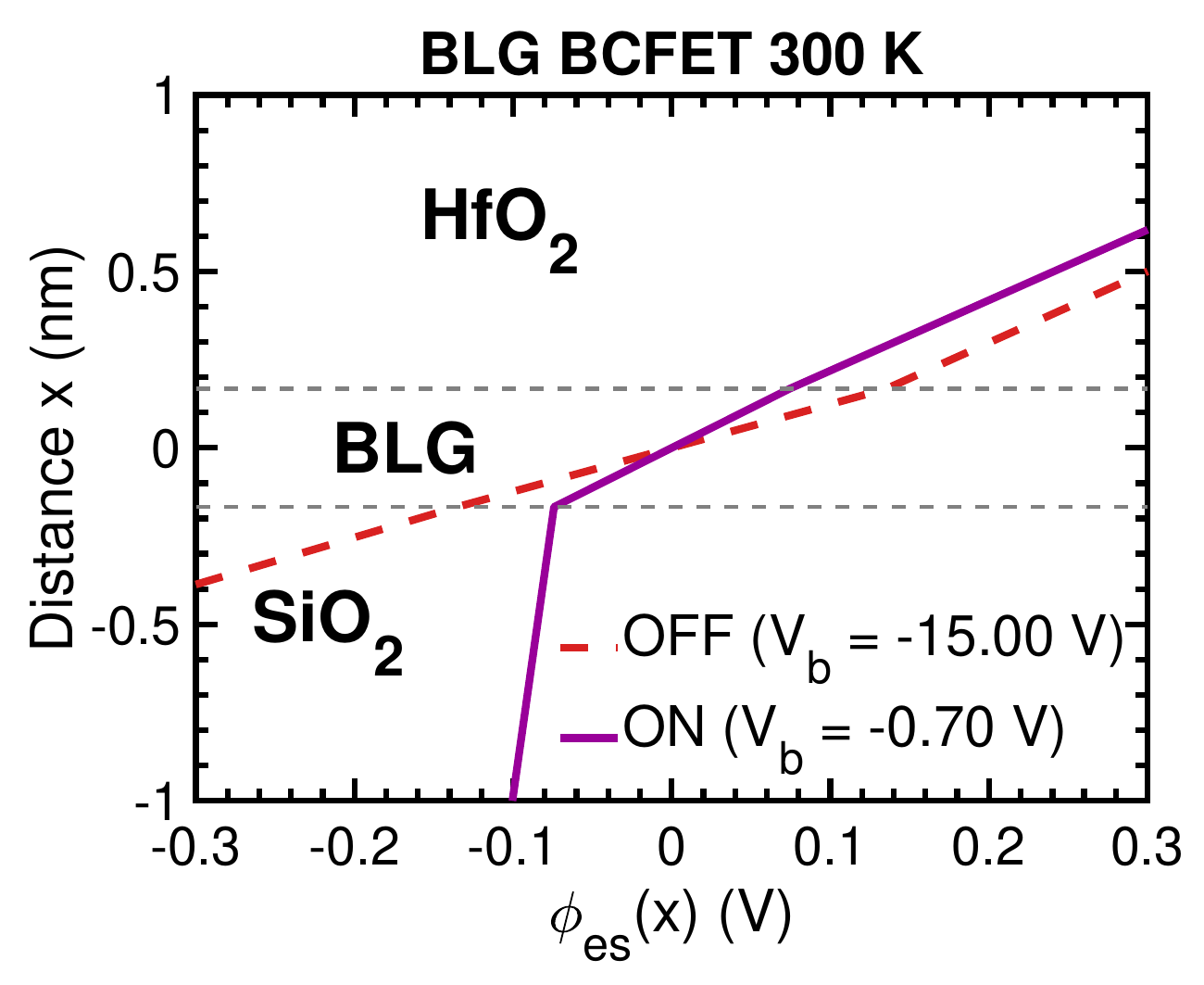}
    \hspace{6pt}
    \panel{0.3885\textwidth}{b}{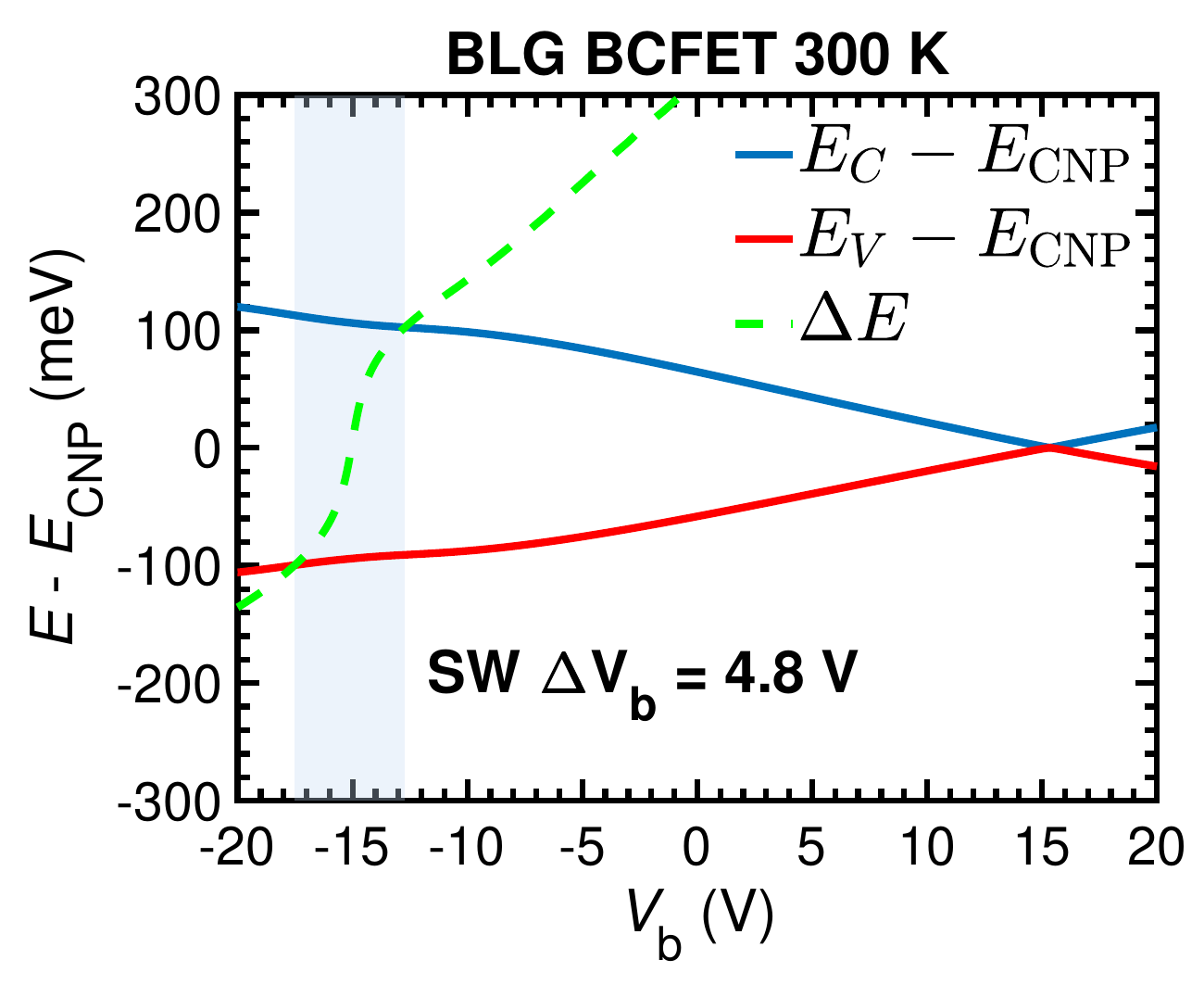}

    \vspace{2pt}

    \panel{0.3885\textwidth}{c}{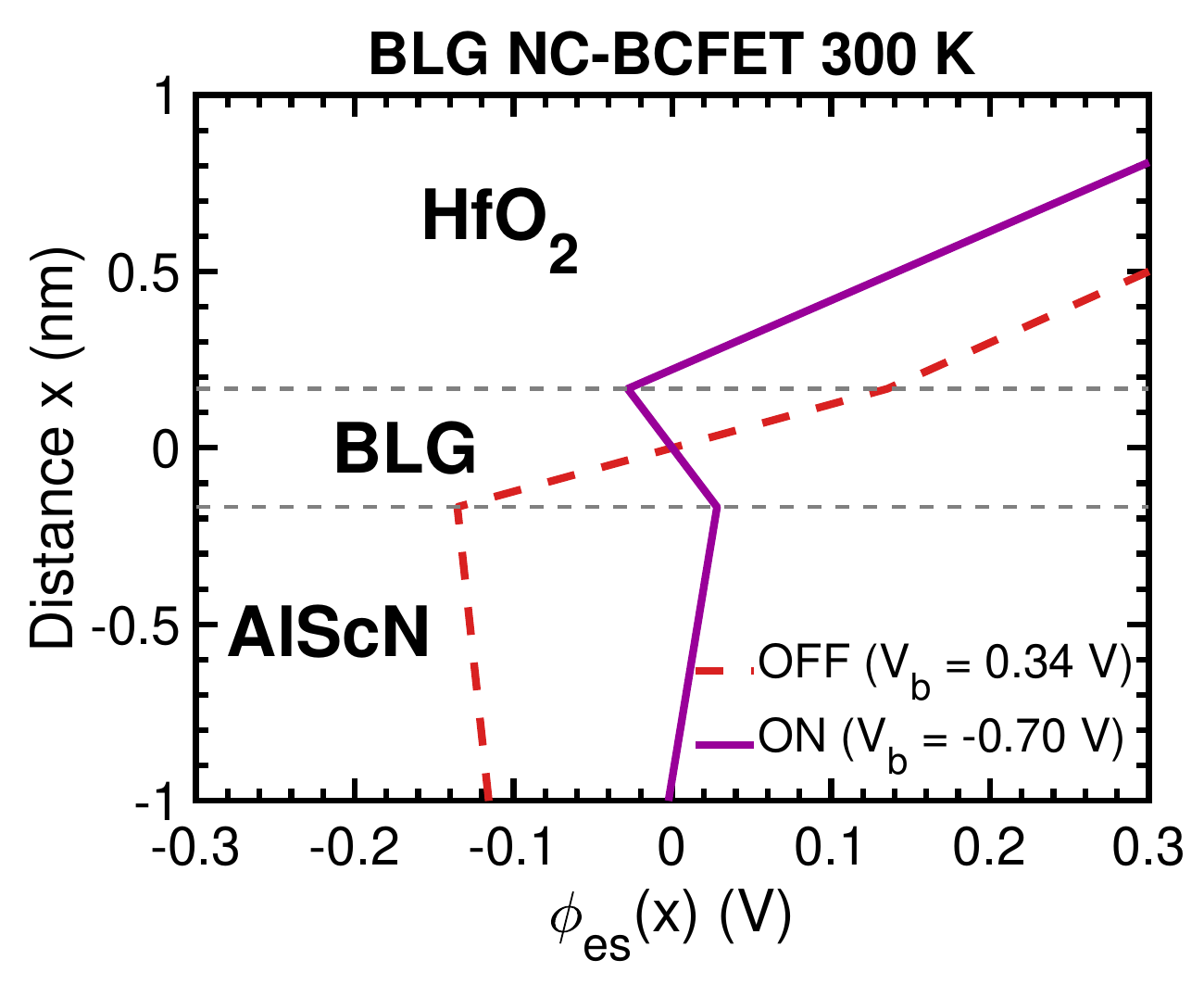}
    \hspace{6pt}
    \panel{0.3885\textwidth}{d}{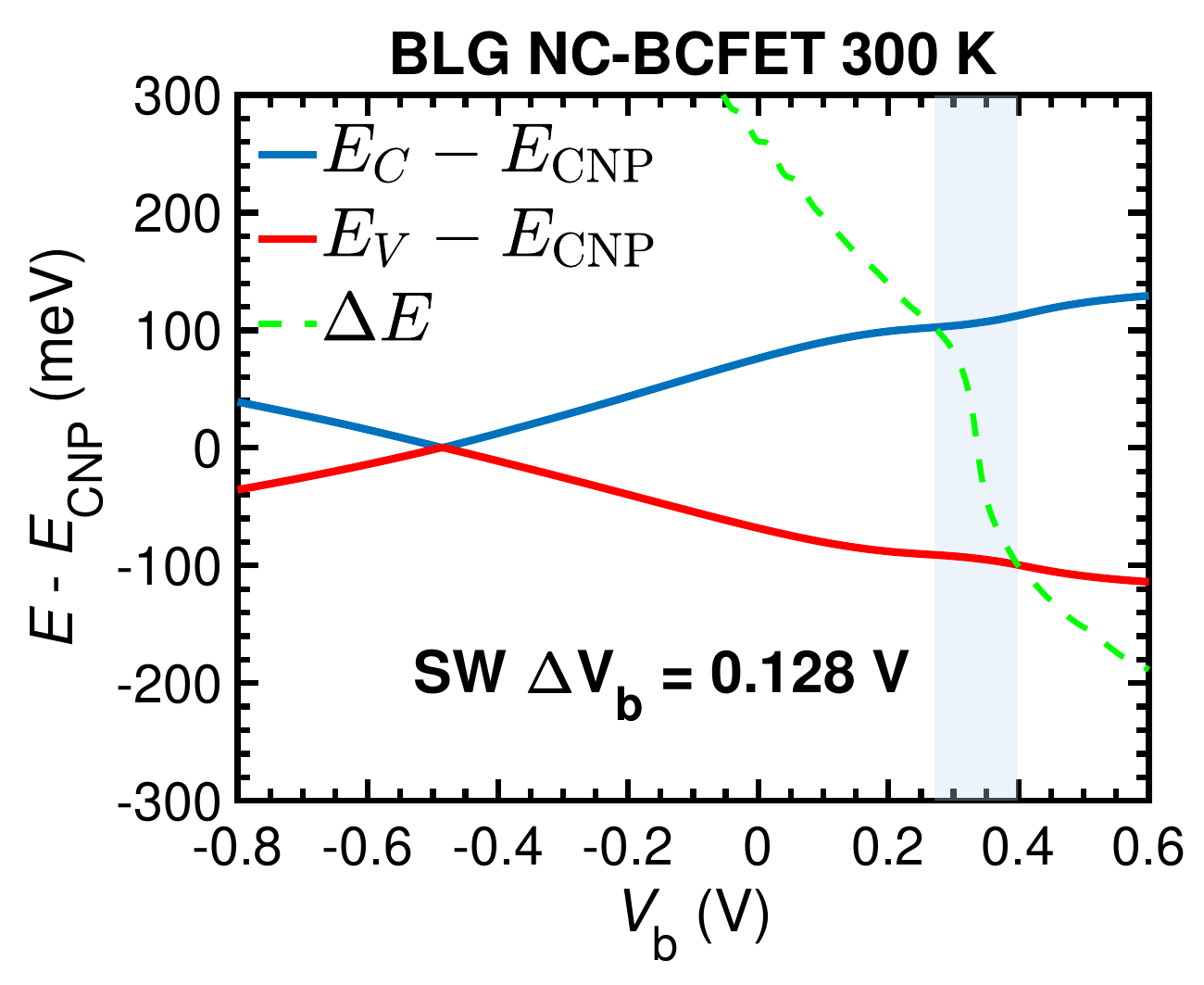}

    \vspace{2pt}

    \panel{0.3885\textwidth}{e}{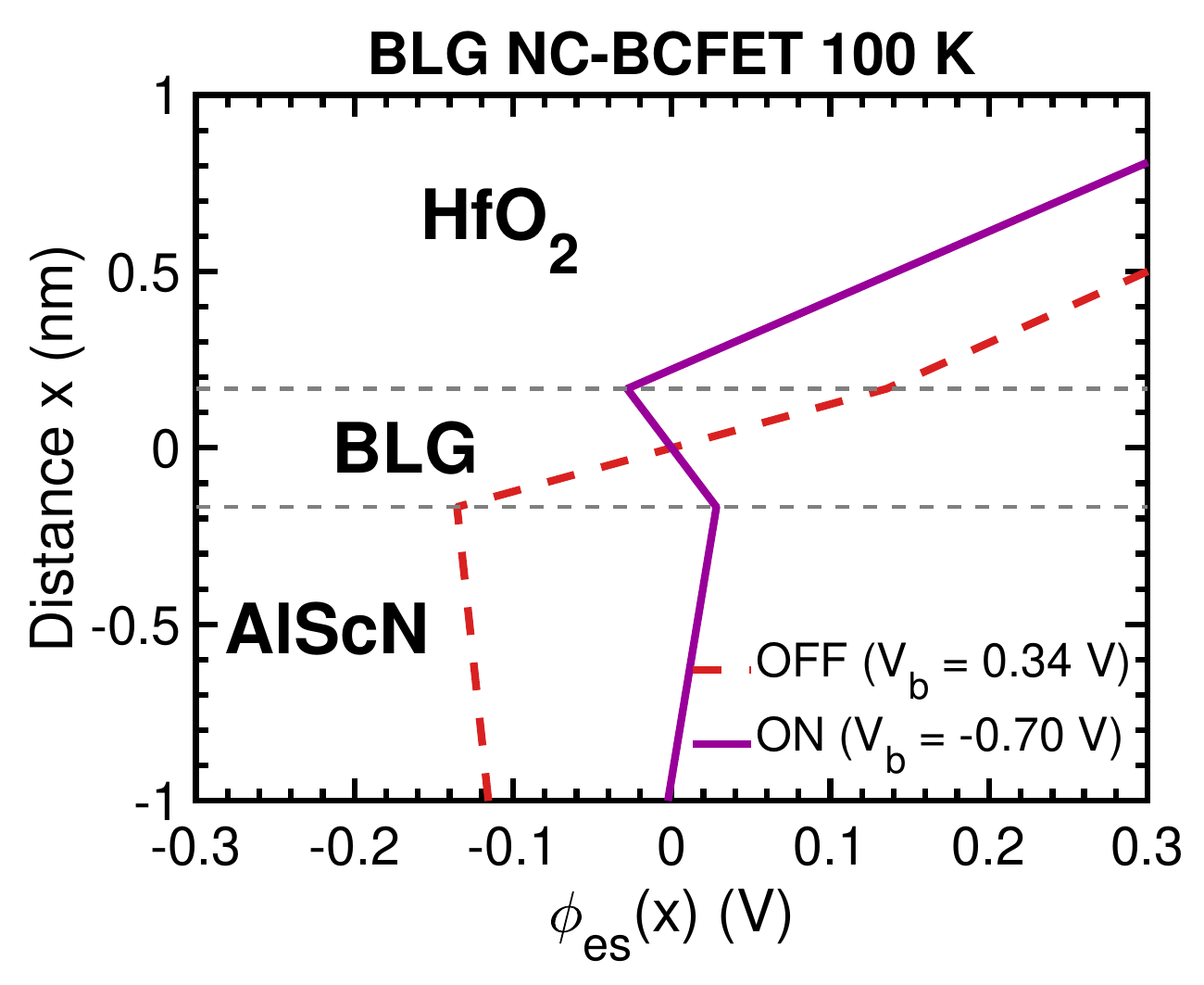}
    \hspace{6pt}
    \panel{0.3885\textwidth}{f}{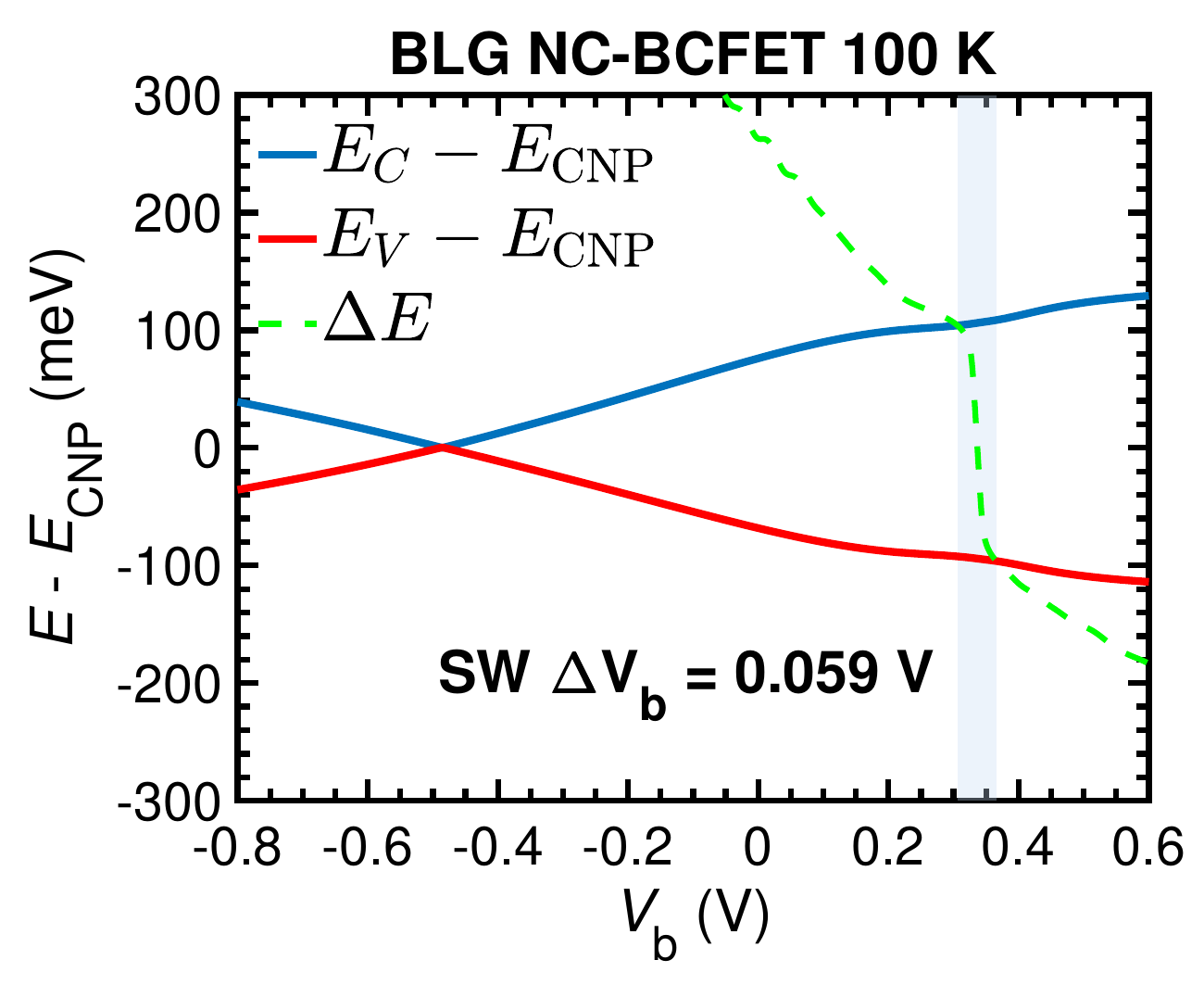}

    \caption{(a) Electrostatic potential profile $\phi_{\mathrm{es}}(x)$ across the HfO$_2$/BLG/SiO$_2$ stack for the BLG BCFET at 300~K. (b) Band edges $E_C - E_{\mathrm{CNP}}$ and $E_V - E_{\mathrm{CNP}}$ with the CNP-referenced Fermi energy $\Delta E = E_F - E_{\mathrm{CNP}} = -U$ versus $V_b$ for the same device. (c,d) As (a,b) for the HfO$_2$/BLG/AlScN BLG NC-BCFET at 300~K, and (e,f) at 100~K. The lightly shaded regions in (b), (d), and (f) mark the switching windows (SW), with $\Delta V_b = 4.79$, 0.128, and 0.059~V, respectively. ON and OFF states are marked on the potential profiles. $\phi_{\mathrm{es}}(x)$ is the Laplace potential of Eqs.~(2.4)--(2.6) of the Supporting Information.}
    \label{fig:2}
\end{figure*}

\indent Figure~\ref{fig:2} shows how the potential profile $\phi_{\mathrm{es}}(x)$ [Figures~\ref{fig:2}(a, c, e)], the conduction and valence band edges $E_C$ and $E_V$, and the band shift $\Delta E$ [Figures~\ref{fig:2}(b, d, f)] vary with gate voltage $V_{b}$. 
We analyze three device configurations: the BLG BCFET at 300~K [Figures~\ref{fig:2}(a, b)], the BLG NC-BCFET at 300~K [Figures~\ref{fig:2}(c,d)], and the BLG NC-BCFET at 100~K [Figures~\ref{fig:2}(e,f)]. 
The potential profile spans the vertical cross-section from the bottom gate electrode ($-L_b$) through the BLG interlayer region to the top gate electrode ($+L_t$).
The potential profiles $\phi_{\mathrm{es}}(x)$ are shown at two characteristic gate voltages: $V_{off,b}$ is the voltage at which $\Delta E = 0$ (charge neutrality), which generally corresponds to the minimum current of the OFF state, while $V_{on,b}$ is a representative gate voltage deep in the ON state ($\Delta E$ located deep in the conduction band). 
The switching-window boundaries $V_{th,n,b}$ and $V_{th,p,b}$ are the threshold voltages for $n-$ and $p-$type conduction, defined as the gate voltages at which $\Delta E$ crosses $E_C - E_{\mathrm{CNP}}$ and $E_V - E_{\mathrm{CNP}}$, respectively. 
The ambipolar switching window is $V_{\mathrm{ASW}} = |V_{th,n,b}-V_{th,p,b}|$, the bottom-gate-voltage interval required for $\Delta E$ to traverse the transport bandgap $E_G = E_C-E_V$. 
Figure~\ref{fig:2}(a) shows that for the dielectric-gated BLG BCFET, the electric field direction, given by the slope of $\phi_{\mathrm{es}}(x)$, is the same in dielectrics and the channel, as expected. 
In contrast, Figures~\ref{fig:2}(c, e) show that for the NC devices the electric field in the channel is opposite in direction to that in the ferroelectric layer, reflecting the negative capacitance of the ferroelectric. 
Moreover, the slope of $\phi_{\mathrm{es}}(x)$ is larger in the BLG channel than in the ferroelectric, reflecting the electric field amplification of the NC effect. 
The switching window (SW) is narrowest for the BLG NC-BCFET with $L_b = 20$~nm and $L_t = 10$~nm at 100~K, with $\Delta V_b = 0.059$~V, compared with 0.128~V at 300~K; the BLG BCFET at 300~K requires $\Delta V_b = 4.79$~V, a factor of 37 wider.

\begin{figure*}[tbp]
    \centering
    \panel{0.3250\textwidth}{a}{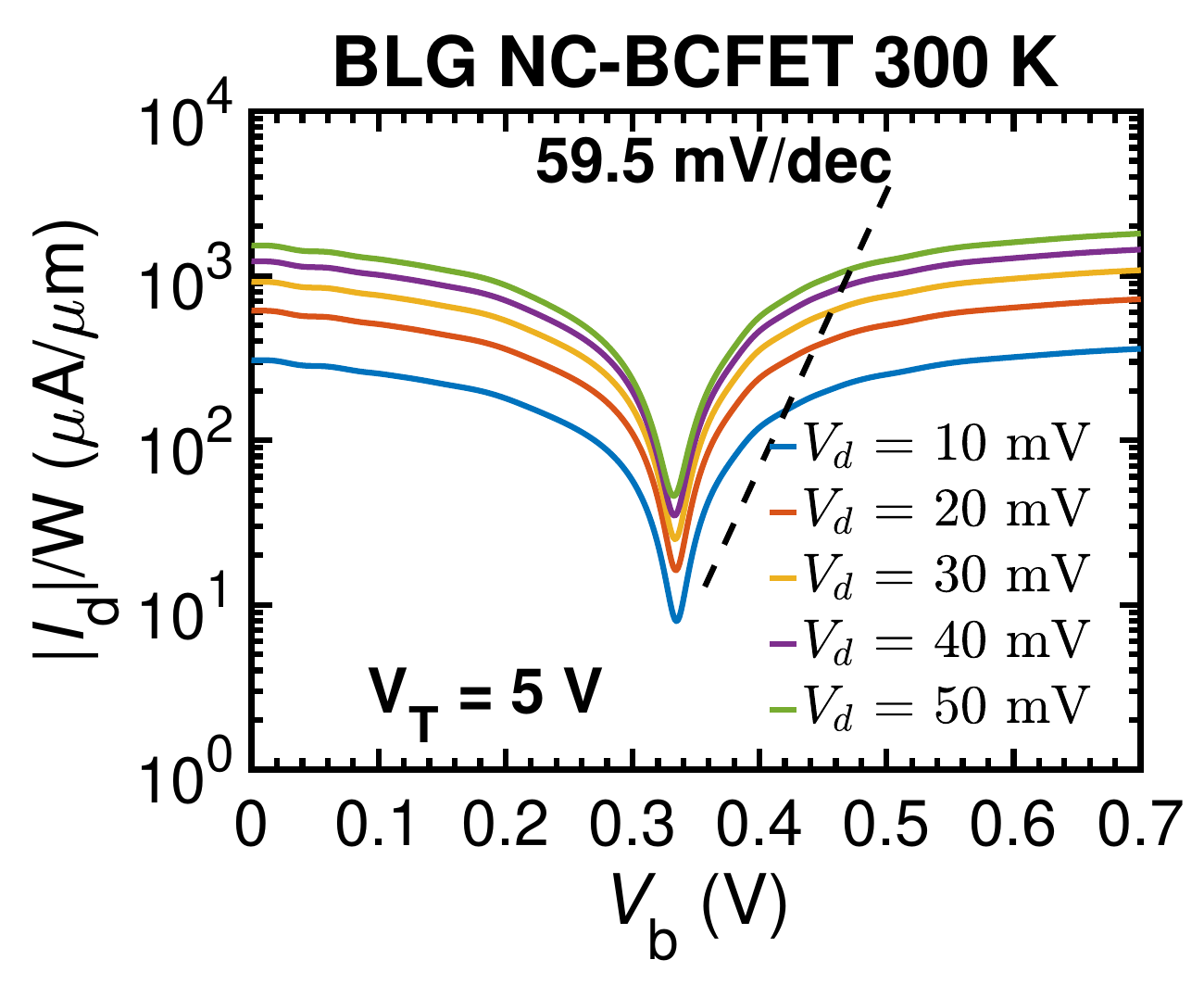}
    \hfill
    \panel{0.3250\textwidth}{b}{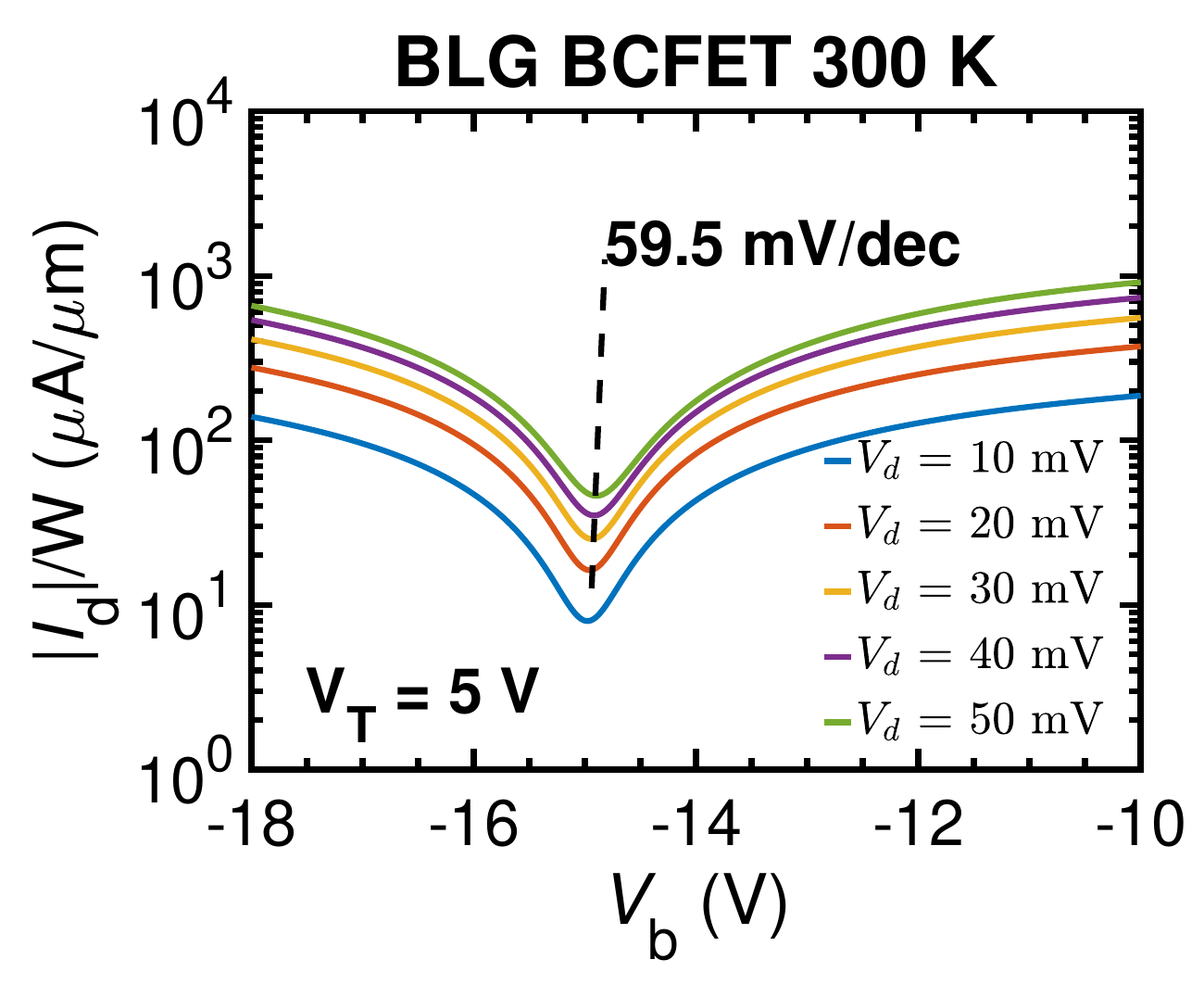}
    \hfill
    \panel{0.3250\textwidth}{c}{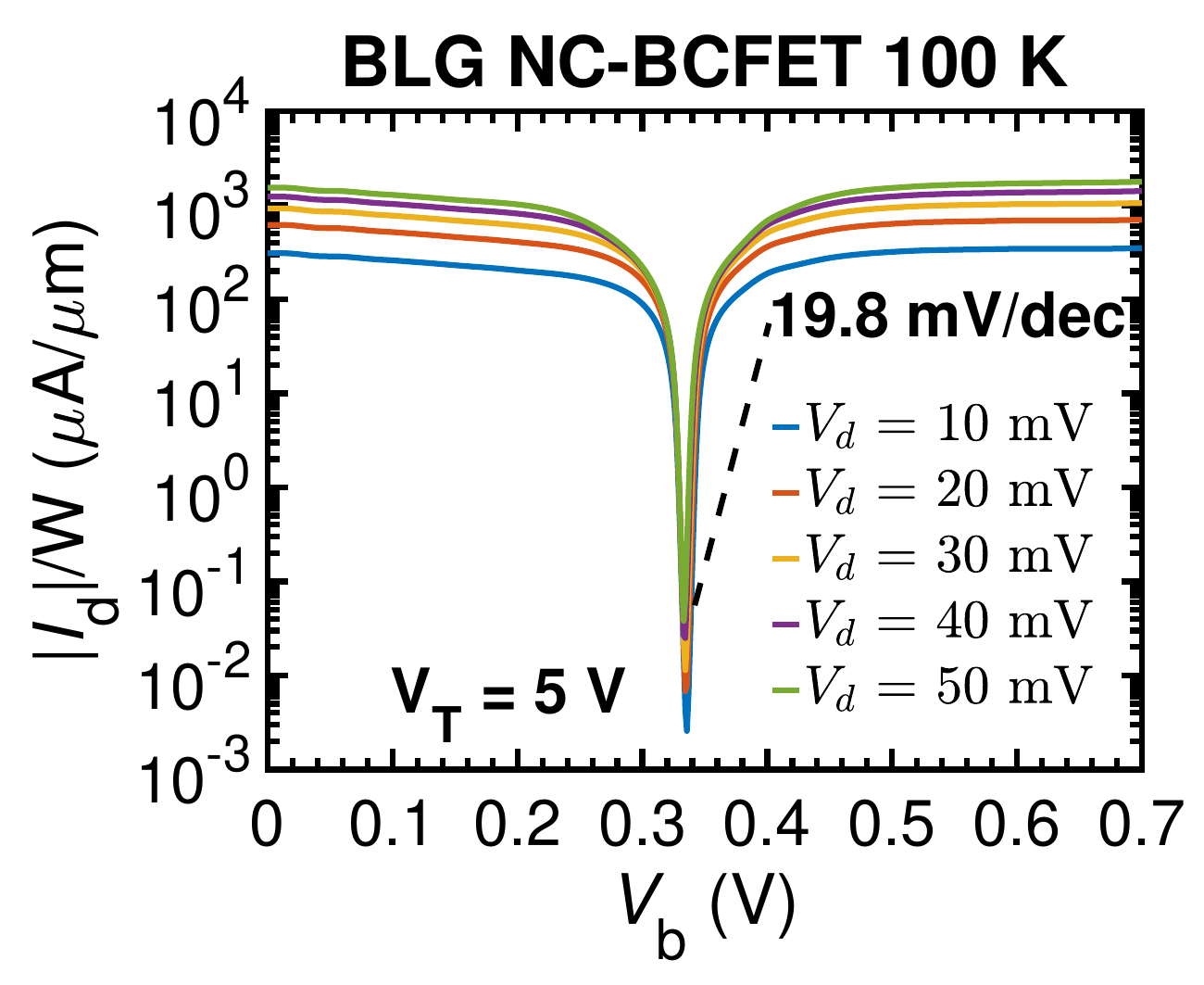}

    \caption{The transfer characteristics (drain current per width $|I_d|/W$ versus $V_b$) at fixed $V_t = 5$~V for $V_d = 10$--50~mV: (a) BLG NC-BCFET at 300~K, (b) BLG BCFET at 300~K, and (c) BLG NC-BCFET at 100~K. The black dashed line in each panel has the slope of the thermionic limit at that temperature, 59.5~mV/dec at 300~K and 19.8~mV/dec at 100~K; it is anchored at the steepest point of the $V_d = 10$~mV trace.}
    \label{fig:3}
\end{figure*}

\indent The narrowing of the switching window follows from the negative capacitance of the ferroelectric. 
In the adopted L-D model, the incremental ferroelectric capacitance is $C_{\mathrm{FE}}=(\varepsilon_0\varepsilon_b+1/\kappa)/t_{\mathrm{FE}}$, where $\kappa=2\alpha_{\mathrm{FE}}+12\beta_{\mathrm{FE}}P_f^2+30\gamma_{\mathrm{FE}}P_f^4$. 
Negative capacitance means $dV_{\mathrm{FE}}/dQ<0$, not necessarily that the absolute $V_{\mathrm{FE}}$ changes sign~\cite{Salahuddin2008,Wong2019}. 
The NC effect allows $\Delta E$ to traverse $E_G$ with a change in gate potential $e\Delta V_b$ that is smaller than the bandgap itself, i.e. $eV_b<E_G$; the chemical potential moves further than the gate potential. 
The ratio $E_G/e\Delta V_b$ is 1.58 at 300~K and 3.40 at 100~K for the BLG NC-BCFET (NC device), compared with 0.042 at 300~K for the BLG BCFET (oxide device).
The transport-gap modulation rates $|dE_G/d\Delta V_b|$ are 173.6~meV/V (BLG NC-BCFET, 300~K) and 4.76~meV/V (BLG BCFET, 300 K), and the corresponding interlayer-asymmetry rates $|d\Delta/dV_b|$ are 320 and 8.8~meV/V; both rates are approximately $36\times$ larger in the BLG NC-BCFET compared to the BLG BCFET. Figures~\ref{fig:2}(b,d,f) show the CNP-referenced band edges and $\Delta E$, whereas Figure~S3 gives the equivalent laboratory frame with the source Fermi level pinned at zero~\cite{Rahman2003}. 
\begin{figure*}[tbp]
    \centering
    \begin{minipage}[t]{0.3250\textwidth}\centering
        \panel{\linewidth}{a}{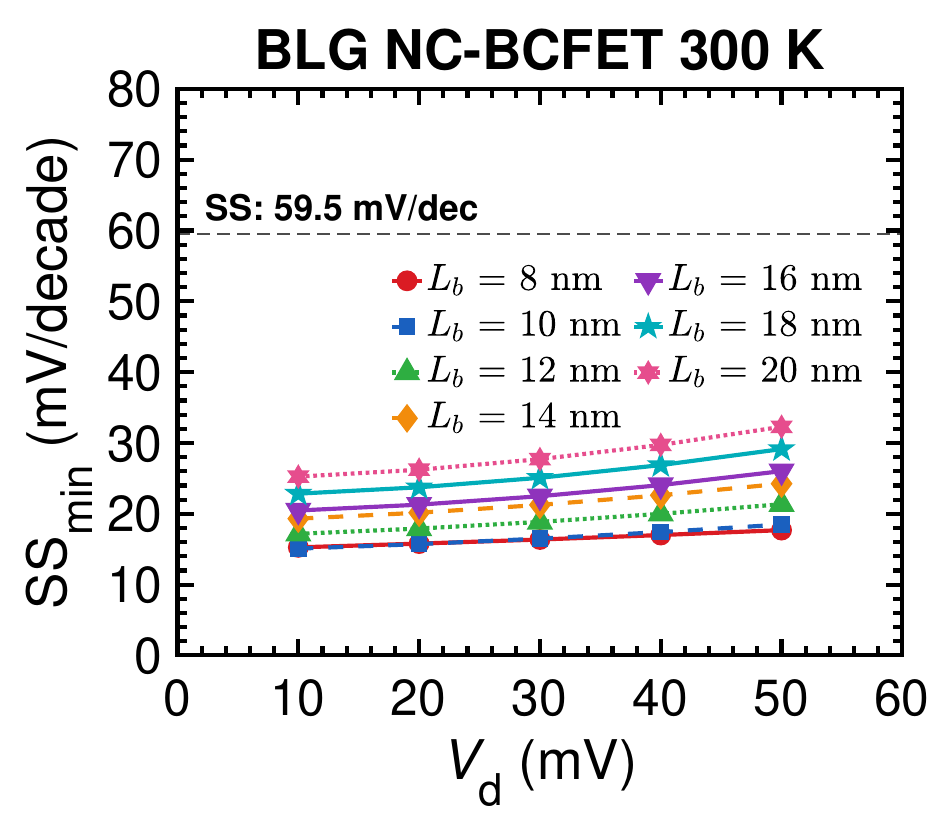}
        \panel{\linewidth}{b}{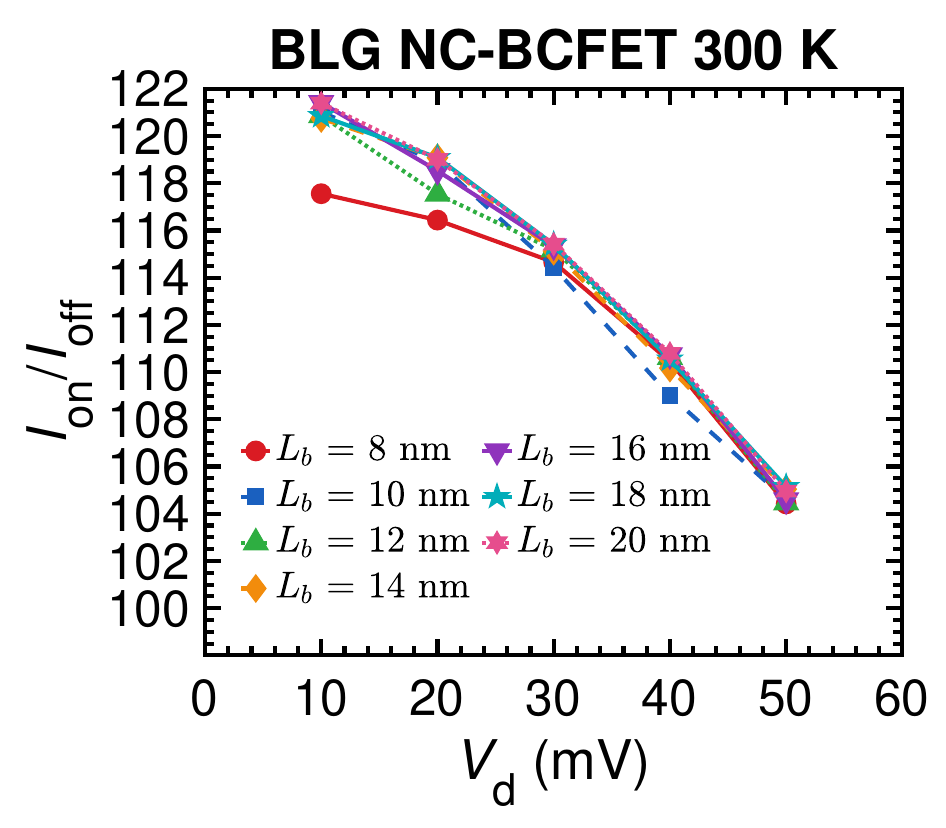}
    \end{minipage}\hfill
    \begin{minipage}[t]{0.3250\textwidth}\centering
        \panel{\linewidth}{c}{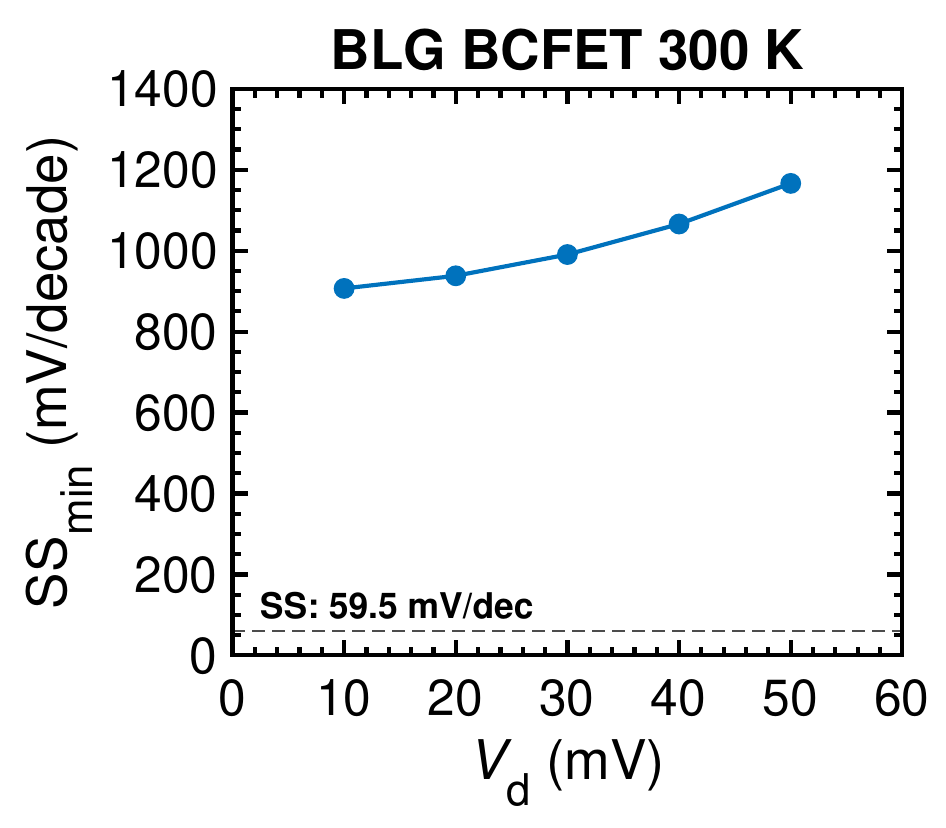}
        \panel{\linewidth}{d}{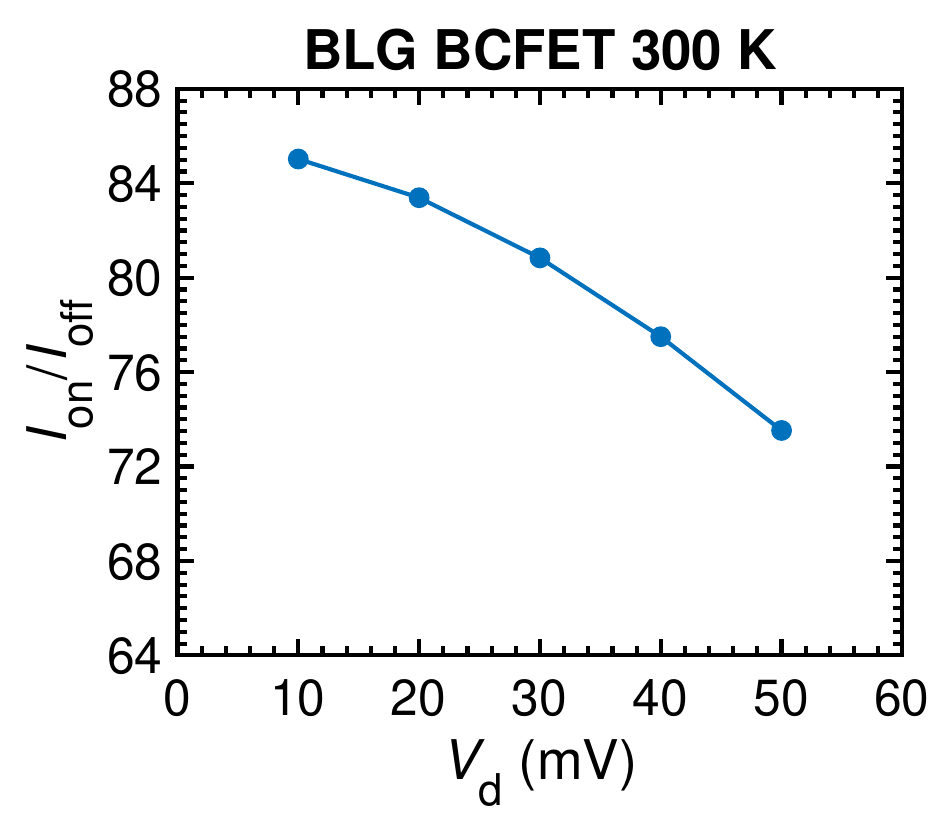}
    \end{minipage}\hfill
    \begin{minipage}[t]{0.3250\textwidth}\centering
        \panel{\linewidth}{e}{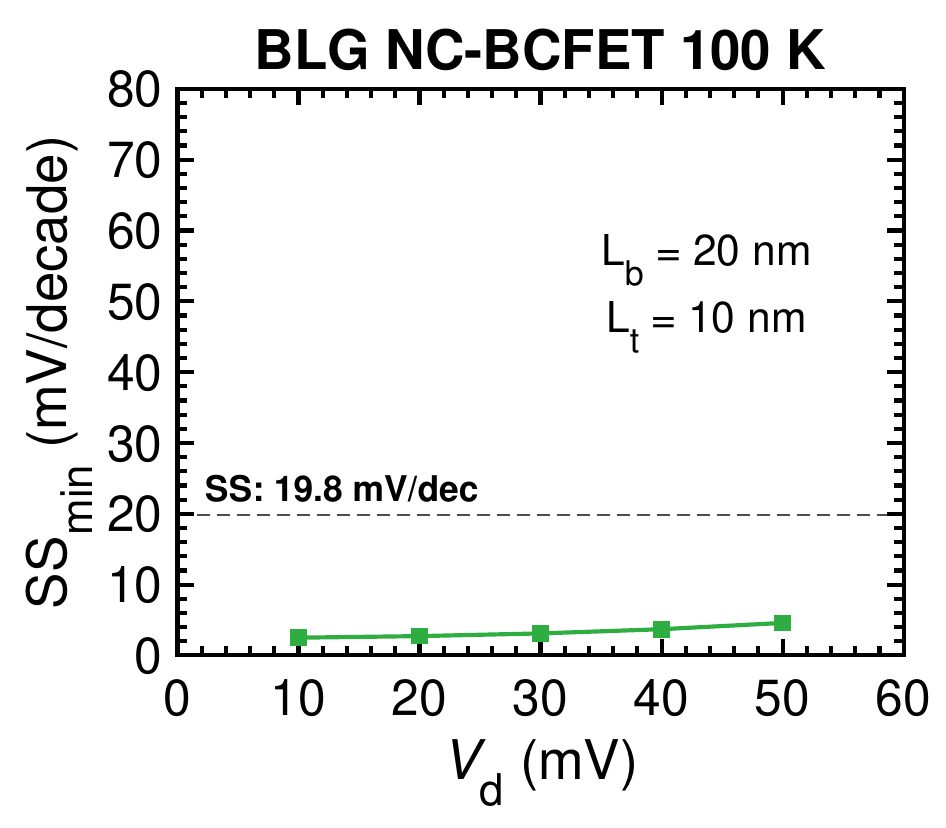}
        \panel{\linewidth}{f}{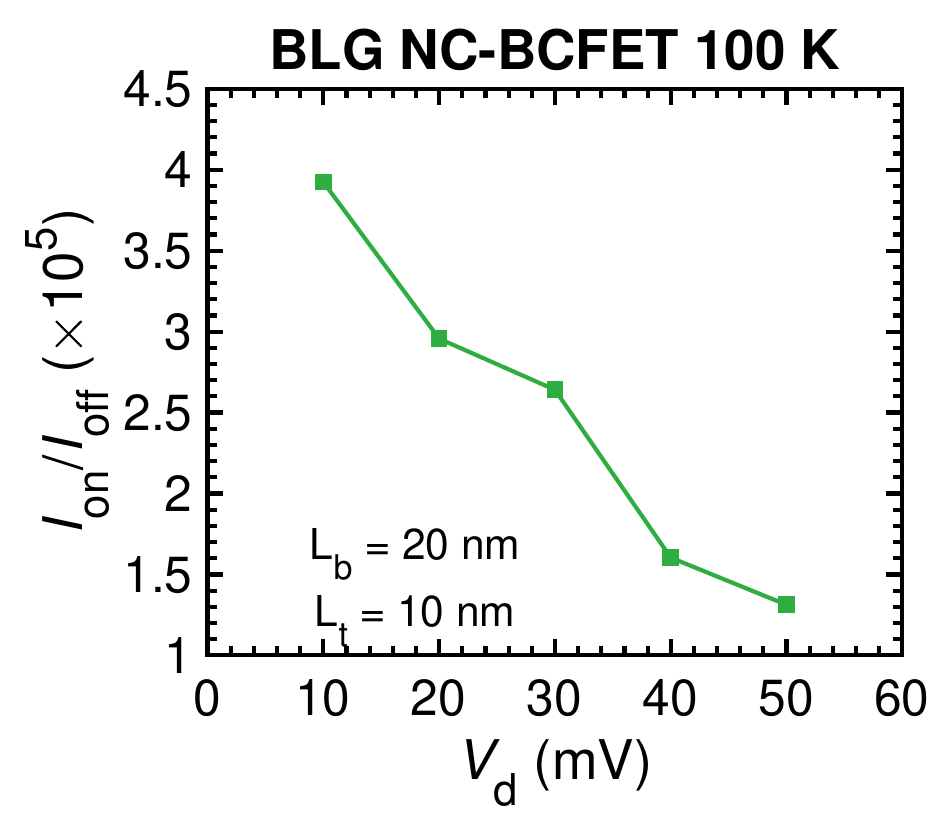}
    \end{minipage}

    \caption{Minimum subthreshold swing $\mathrm{SS_{min}}$ (top row) and on/off current ratio $I_{\mathrm{on}}/I_{\mathrm{off}}$ (bottom row) versus drain bias $V_d$: (a,b) BLG NC-BCFET at 300~K for ferroelectric thickness $L_b = 8$--20~nm, (c,d) the reference BLG BCFET at 300~K, and (e,f) the $L_b = 20$~nm BLG NC-BCFET at 100~K. The dashed line in (a), (c) and (e) is the thermionic limit at that temperature, 59.5~mV/dec at 300~K and 19.8~mV/dec at 100~K. Panels (c,d) are shown in blue and panels (e,f) in green. Device-specific on/off voltage windows are stated in the Supporting Information.}
    \label{fig:4}
\end{figure*}

\indent At lower temperature (100~K), the SW narrows further, to about $81\times$ smaller than the oxide device at 300~K. 
Reduced thermal smearing sharpens the Fermi--Dirac transition, and in combination with NC amplification, yields the steepest subthreshold slope for the BLG NC-BCFET at 100~K (see below). 
The thermal broadening window is $\pm 2k_BT$ from the Fermi level; at 100~K ($k_BT \approx 8.6$~meV) it is approximately three times narrower than at 300~K ($k_BT \approx 25.9$~meV). 
This temperature dependence is consistent with the cryogenic BLG FET results of Icking \textit{et al.}~\cite{Icking2024}, who observed ultrasteep subthreshold slopes in BLG devices at low temperatures. Combining NC amplification with reduced thermal broadening provides the physical mechanism for the abrupt band-edge transitions that yield ultrasteep switching~\cite{Icking2024}.

\indent The amplification also shows in the gate voltage needed to open the gap: the BLG NC-BCFET device reaches a gap of approximately 261~meV at $V_b=0.5$~V, whereas its oxide counterpart requires $V_b=-20$~V for a comparable 253~meV gap. 
At the same bias $V_b=-0.7$~V and $V_d=10$~mV, the width-normalized ON currents are 720.8 A/m for the 20 nm FE device and 336.6 A/m for the BLG BCFET, an enhancement of 2.14.


\indent Figure~\ref{fig:3} plots the transfer characteristics of the modeled devices. 
At 300 K the BLG NC-BCFET (Figure~\ref{fig:3}a) and the BLG BCFET (Figure~\ref{fig:3}b) show V-shaped ambipolar transfer characteristics with modest on-off ratios of order 10$^{2}$, reflecting the small bandgap of BLG. 
The black dashed lines indicate the thermionic limit $SS=\ln(10)k_BT/q$. 
The BLG NC-BCFET switches more steeply than the thermionic limit, whereas the BLG BCFET is far shallower. Figure~\ref{fig:3}c shows the BLG NC-BCFET at a temperature of 100 K. 
The on-off ratio increases to above $10^{5}$, and the transfer characteristics remain steeper than the thermionic limit over a wider range of currents.

\indent Figure~\ref{fig:4} shows the lowest subthreshold swings $\mathrm{SS_{min}}$ and on-off ratios $I_{\mathrm{on}}/I_{\mathrm{off}}$ extracted from the transfer characteristics shown in Figure~\ref{fig:3}. 
Figures~\ref{fig:4}a,b present the NC-BCFET at 300~K. 
For a ferroelectric thickness $L_b=20$~nm at low drain voltage ($V_d=10$~mV), $\mathrm{SS_{min}}$ is 25~mV/dec, well below the thermionic limit of 60 mV/dec, and $I_{\mathrm{on}}/I_{\mathrm{off}}$ is 121.
$\mathrm{SS_{min}}$ decreases further to 15 mV/dec (4 times lower than the thermionic limit) when the ferroelectric thickness is reduced to 10 nm, with only a small decrease in $I_{\mathrm{on}}/I_{\mathrm{off}}$.

Figures~\ref{fig:4}c,d show the reference dielectric-gated BLG BCFET at 300~K. 
At $V_d=10$~mV, $\mathrm{SS_{min}}$ is 906~mV/dec, 15 times the thermionic limit, and the on-off ratio is lower at 85. 
In our BLG BCFET, $V_b$ is swept through the 20~nm SiO$_2$ bottom dielectric, whose equivalent oxide thickness (EOT) is therefore also 20~nm, while the HfO$_2$ top gate (EOT $\approx$ 6.5~nm) is held fixed. 
While the gate coupling could be improved, 
even in an idealised 1~nm EOT stack the lowest SS is $\sim$200--275~mV/dec, and measured BLG FETs give about 550~mV/dec at room temperature, so dielectric-gated BLG BCFETs are unlikely to approach the thermionic limit at room temperature regardless of geometry~\cite{Majumdar2010,Xia2010}. Figures~\ref{fig:4}e,f show that reducing the temperature to 100 K can improve $\mathrm{SS_{min}}$ to 2.5~mV/dec, 7.9 times below the 100~K thermionic limit of 19.8~mV/dec and $I_{\mathrm{on}}/I_{\mathrm{off}}$ to $3.9\times10^{5}$ at $V_d=10$~mV with $L_b=20$~nm.


\section{\label{sec:conclusion}Conclusion}

\indent In summary, we have developed a self-consistent device model of a bilayer graphene negative-capacitance bandgap-change field-effect transistor, coupling a four-band tight-binding Hamiltonian with GW-corrected screening, a quasistatic Landau--Devonshire description of the Al$_{0.55}$Sc$_{0.45}$N ferroelectric, and ballistic top-of-the-barrier transport. The negative capacitance of the gate stack amplifies the electric field in the channel, so that the chemical potential traverses the transport gap with a gate swing smaller than the gap itself: $E_G/e\Delta V_b=1.58$ at 300~K and 3.40 at 100~K, and the switching window is compressed 37-fold relative to the dielectric-gated control. This field amplification, rather than a reduced body factor alone, is the origin of the steep switching. The device operates in the negative-capacitance regime for ferroelectric thicknesses of 8--20~nm and switches below the thermionic limit throughout that range, with an optimized $\mathrm{SS_{min}}=15$~mV/dec for $L_b=10$~nm, four times below the Boltzmann limit. At 100 K, reduced thermal broadening compounds the amplification, and the device reaches $\mathrm{SS_{min}}=2.5$~mV/dec for $L_b=20$~nm with $I_{\mathrm{on}}/I_{\mathrm{off}}=3.9\times10^{5}$.

\indent The room-temperature on/off ratio is limited to about $10^{2}$ by the largest field-tunable bandgap attainable in bilayer graphene, approximately 260~meV. This limitation notwithstanding, the switching mechanism does not depend on any property specific to bilayer graphene beyond an electrically tunable gap, and the framework applies without modification to any bandgap-change channel. Two-dimensional materials with substantially larger tunable bandgaps have been reported both theoretically and experimentally~\cite{Ramasubramaniam2011,Deng2017}; in such channels the same negative-capacitance gate stack would deliver deep subthermionic switching together with high on/off ratios at room temperature. Extending the present framework to these materials is the immediate next step of the work presented here.

\section{\label{sec:methods}Methods}

\indent All reported device characteristics come from the stored self-consistent numerical solutions. The BLG band structure uses the \textit{ab initio} GW screening parametrization of Gava \textit{et al.}~\cite{Gava2009}; the Al$_{0.55}$Sc$_{0.45}$N response is the quasistatic equilibrium limit of a sixth-order Landau--Devonshire model~\cite{Gu2024,Salahuddin2008,Wong2019}; and the saved finite-drain current is evaluated with the Landauer--B\"uttiker top-of-the-barrier formulation~\cite{Rahman2003,Lundstrom2017}. Quantum capacitance enters through the DOS-weighted charge $n(U)$ rather than as a separate lumped capacitor. In the FE calculation, the stored configuration uses the incremental $C_{\mathrm{FE}}(P_f)$ in the large-signal Laplace numerator and the geometric background capacitance in $C_{\Sigma}$; this hybrid coupling is an explicit model assumption. The common-mode channel shift is carried by $U$ in the Fermi functions, with $\Delta E=-U$, and the resulting charge feeds back into the polarization, GW screening coefficient, and interlayer asymmetry at every gate bias. 

\noindent\textbf{Supporting Information.} Device structure, electrostatic and ferroelectric models, self-consistent solution procedure, transport calculations, subthreshold-swing extraction, and ferroelectric-thickness optimization.

\section*{Acknowledgements}
R.B.A., M.S.F., and N.M. acknowledge financial support from the IITB-Monash Research Academy. This research was undertaken with the assistance of resources from the National Computational Infrastructure (NCI Australia), an NCRIS enabled capability supported by the Australian Government. The author BM acknowledges insightful discussions with V. Deshpande. The author BM acknowledges funding from the Department of Science and Technology (DST), Government of India, under the National Quantum Mission {through Grant no. DST/QTC/NQM/QMD/2024/4} and the Inani Chair Professorship fund{, through Grant No. DO/2024-INAN/001-001}. The author BM also acknowledges the Annual Travel Grant from the Quantum Materials and Devices hub (QMD hub through Grant no. QMDF/ITG/2025-26/4), which facilitated a visit to Monash University.

\appendix

\section{\label{app:tb}Four-band Hamiltonian and field-tunable bandgap}

The appendices summarize the analytic relations behind the numerical model; the full formulation is given in the Supporting Information.

Near the $K_+$ point, in the basis $(A_1,B_1,A_2,B_2)$ with layer~1 adjacent to the bottom gate and $\mathbf{k}$ measured from $K_+$, BLG is described by the four-band Hamiltonian~\cite{McCann2013}
\begin{equation}
H=\begin{pmatrix}
-\Delta/2 & v\pi^{\dagger} & -v_4\pi^{\dagger} & v_3\pi\\
v\pi & -\Delta/2 & \gamma_1 & -v_4\pi^{\dagger}\\
-v_4\pi & \gamma_1 & \Delta/2 & v\pi^{\dagger}\\
v_3\pi^{\dagger} & -v_4\pi & v\pi & \Delta/2
\end{pmatrix},
\label{eq:app-H}
\end{equation}
where $\pi=\hbar(k_x+ik_y)$, $v=\sqrt{3}a\gamma_0/2\hbar$, $v_{3,4}=\sqrt{3}a\gamma_{3,4}/2\hbar$, and $\Delta=\epsilon_2-\epsilon_1$ is the interlayer asymmetry, with $\gamma_0=3.16$, $\gamma_1=0.39$, $\gamma_3=0.38$, and $\gamma_4=0.14$~eV. Neglecting $\gamma_3$ and $\gamma_4$, the low-energy bands take a Mexican-hat shape with the reduced-model gap~\cite{McCann2006,McCann2013}
\begin{align}
E_{\mathrm{gap}}^{(0)}&=\frac{|\Delta|\gamma_1}{\sqrt{\Delta^{2}+\gamma_1^{2}}},\label{eq:app-gap}\\
\frac{dE_{\mathrm{gap}}^{(0)}}{d|\Delta|}&=\frac{\gamma_1^{3}}{\left(\Delta^{2}+\gamma_1^{2}\right)^{3/2}}.\label{eq:app-gapslope}
\end{align}
The reduced-model gap follows $|\Delta|$ for $|\Delta|\ll\gamma_1$ and saturates at $\gamma_1$ for $|\Delta|\gg\gamma_1$; at the asymmetries of the switching windows, $|\Delta|\approx0.26$--0.30~eV, the slope in Eq.~(\ref{eq:app-gapslope}) is 0.50--0.58. The nominal gap values in Table~S2, the quoted nominal-gap modulation rates, and the in-gap thermal-activation floor use $E_{\mathrm{gap}}^{(0)}$. In contrast, the band edges $E_C$ and $E_V$, density of states, mode count, and threshold crossings are obtained from the full spectrum of Eq.~(\ref{eq:app-H}), for which the transport gap is $E_G=E_C-E_V$. Thus $E_{\mathrm{gap}}^{(0)}$ is an analytic gap proxy and need not equal $E_G$ when $\gamma_3$ and $\gamma_4$ are retained.

\section{\label{app:asym}Ferroelectric amplification of the interlayer asymmetry}

According to  the GW parametrization of Gava \textit{et al.}~\cite{Gava2009} is $\Delta=\alpha_{\mathrm{GW}}(n)\,(n_2-n_1)$, where $n_1$ and $n_2$ are the sheet densities induced by the bottom and top gates. Gauss's law gives
\begin{align}
en_2&=\bar C_t\left(V_t+\frac{\Delta}{2e}\right),\nonumber\\
en_1&=
\begin{cases}
\bar C_b\left(V_b-\dfrac{\Delta}{2e}\right),&\text{BCFET},\\
\varepsilon_0\varepsilon_bE_{\mathrm{geo}}+P_f,&\text{NC-BCFET},
\end{cases}
\label{eq:app-gauss}
\end{align}
where $\bar C_t=\varepsilon_0\varepsilon_{r,t}/L_t$, $\bar C_b=\varepsilon_0\varepsilon_{r,b}/L_b$, and $E_{\mathrm{geo}}=(V_b +\Delta/2e)/t_{\mathrm{FE}}$. For the NC-BCFET, $P_f$ is set by $E_{\mathrm{geo}}=2\alpha_{\mathrm{FE}}P_f+4\beta_{\mathrm{FE}}P_f^{3}+6\gamma_{\mathrm{FE}}P_f^{5}$. Locally linearizing about $P_f=0$ gives
\begin{equation}
C_{\mathrm{FE}}\simeq\frac{1}{t_{\mathrm{FE}}}\left(\varepsilon_0\varepsilon_b+\frac{1}{2\alpha_{\mathrm{FE}}}\right),
\label{eq:app-cfe}
\end{equation}
the $P_f\to0$ differential capacitance in the main text ($-0.0545$~F/m$^2$ for $t_{\mathrm{FE}}=20$~nm), so locally $en_1\simeq C_{\mathrm{FE}}(V_b-\Delta/2e)$. The negative-capacitance branch is physically accessible here only as a series-stabilized state of the complete gate--channel stack~\cite{Salahuddin2008,Wong2019}. Holding the capacitances and $\alpha_{\mathrm{GW}}$ fixed in this local approximation, and defining $C_\alpha\equiv2e^{2}/\alpha_{\mathrm{GW}}$ when $\alpha_{\mathrm{GW}}$ is expressed in J\,m$^2$ (equivalently, $C_\alpha=2e/\alpha_{\mathrm{GW}}$ when it is expressed in eV\,m$^2$), Eq.~(\ref{eq:app-gauss}) gives
\begin{align}
\Delta&\simeq
\begin{cases}
\dfrac{2e\,(\bar C_bV_b-\bar C_tV_t)}{C_\alpha+\bar C_t+\bar C_b},&\text{BCFET},\\[6pt]
\dfrac{2e\,(C_{\mathrm{FE}}V_b-\bar C_tV_t)}{C_\alpha+\bar C_t+C_{\mathrm{FE}}},&\text{NC-BCFET},
\end{cases}
\label{eq:app-delta}\\
\frac{d\Delta}{dV_b}&\simeq
\begin{cases}
\dfrac{2e\,\bar C_b}{C_\alpha+\bar C_t+\bar C_b},&\text{BCFET},\\[6pt]
\dfrac{2e\,C_{\mathrm{FE}}}{C_\alpha+\bar C_t+C_{\mathrm{FE}}},&\text{NC-BCFET}.
\end{cases}
\label{eq:app-ddelta}
\end{align}
The GW screening capacitance, $C_\alpha\approx0.37$~F/m$^2$ in the switching windows, is much larger than $\bar C_t$, $\bar C_b$, and $|C_{\mathrm{FE}}|$, so the local response of the asymmetry is approximately proportional to the corresponding bottom-gate capacitance. The ferroelectric replaces $\bar C_b$ by $C_{\mathrm{FE}}$, which is negative and, for the same thickness and background permittivity, larger in magnitude by $1/(2|\alpha_{\mathrm{FE}}|\varepsilon_0\varepsilon_b)-1\approx32$. The response of $\Delta$ therefore reverses sign and is enhanced by a factor of $\approx37$ at fixed $\alpha_{\mathrm{GW}}$, accounting for the $\approx36$-fold larger modulation rates of the NC-BCFET in the main text; the small difference comes from the variation of $\alpha_{\mathrm{GW}}(n)$ and the nonlinear ferroelectric response across the switching windows. The nominal gap proxy follows through Eq.~(\ref{eq:app-gapslope}). This is the electric-field amplification proposed for the NC-TQFET~\cite{fuhrer2021nctqfet}. Within the same local approximation, Eq.~(\ref{eq:app-delta}) estimates that $\Delta$ vanishes at $V_b=\bar C_tV_t/\bar C_b$ for the BCFET and at $V_b=\bar C_tV_t/C_{\mathrm{FE}}$ for the NC-BCFET, namely $+15.3$~V and $-0.49$~V, respectively, where the band edges in Figures~\ref{fig:2}(b) and \ref{fig:2}(d) meet.

\section{\label{app:ss}Subthreshold swing of a bandgap-change FET}

\subsection{Nondegenerate top-of-the-barrier current}

In the ToB model the drain current is~\cite{Rahman2003,Lundstrom2017}
\begin{equation}
I_{\mathrm{DS}}=\frac{4q}{h}\int M(E)\left[f_S(E)-f_D(E)\right]dE,
\label{eq:app-current}
\end{equation}
where $f_S(E)=f(E+U)$ and $f_D(E)=f(E+U+qV_d)$ are the source and drain occupations, $f$ is the Fermi function, the grounded source sets the zero of energy, $U=-\Delta E$, and $M(E)$ is the number of right-moving modes per spin and per valley of the unshifted bands, with energies measured from $E_{\mathrm{CNP}}$ as in Figures~\ref{fig:2}(b,d,f). When the relevant source and drain occupations are nondegenerate, with $E_C-E_F\gg k_BT$, $E_F-E_V\gg k_BT$, and, for the hole branch at positive $V_d$, $E_F-E_V-qV_d\gg k_BT$, the electron and hole currents become
\begin{align}
I_n&=\frac{4q}{h}\left(1-e^{-qV_d/k_BT}\right)e^{-(E_C-E_F)/k_BT}J_C,\label{eq:app-In}\\
I_p&=\frac{4q}{h}\left(e^{qV_d/k_BT}-1\right)e^{-(E_F-E_V)/k_BT}J_V,\label{eq:app-Ip}
\end{align}
with $E_C-E_F=(E_C-E_{\mathrm{CNP}})-\Delta E$, $E_F-E_V=\Delta E-(E_V-E_{\mathrm{CNP}})$, and
\begin{align}
J_C&=\int_0^{\infty}M(E_C-E_{\mathrm{CNP}}+x)\,e^{-x/k_BT}\,dx,\nonumber\\
J_V&=\int_0^{\infty}M(E_V-E_{\mathrm{CNP}}-x)\,e^{-x/k_BT}\,dx.
\label{eq:app-J}
\end{align}
The integrals $J_C$ and $J_V$ inherit a gate dependence through the band structure; the analytic estimate below neglects their logarithmic derivatives with respect to $V_b$.

\subsection{Subthreshold swing}

On a flank dominated by one carrier type, and neglecting $d\ln J_{C,V}/dV_b$, Eqs.~(\ref{eq:app-In}) and (\ref{eq:app-Ip}) give
\begin{equation}
SS\simeq\ln(10)\,\frac{k_BT}{q}\left|\frac{1}{q}\frac{dE_{\mathrm{B}}}{dV_b}\right|^{-1},
\label{eq:app-ss}
\end{equation}
with $E_{\mathrm{B},n}=E_C-E_F$ and $E_{\mathrm{B},p}=E_F-E_V$, whose derivatives are
\begin{equation}
\begin{aligned}
\frac{dE_{\mathrm{B},n}}{dV_b}&=\frac{d(E_C-E_{\mathrm{CNP}})}{dV_b}-\frac{d\Delta E}{dV_b},\\
\frac{dE_{\mathrm{B},p}}{dV_b}&=\frac{d\Delta E}{dV_b}-\frac{d(E_V-E_{\mathrm{CNP}})}{dV_b}.
\end{aligned}
\label{eq:app-decomp}
\end{equation}
In a conventional FET the band edge is fixed, $q^{-1}|d\Delta E/dV_b|=1/m$, and Eq.~(\ref{eq:app-ss}) reduces to Eq.~(\ref{Eq:eq1}); in a BCFET the field-driven motion of the band edge (Appendix~\ref{app:asym}) enters as an additional term.

Define the cross-bias energy span of the switching window as $\Delta E_{\mathrm{SW}}\equiv|\Delta E(V_{th,n,b})-\Delta E(V_{th,p,b})|$. Because the band edges vary with $V_b$, this span need not equal the instantaneous full-spectrum gap $E_G$ at either endpoint. For an approximately symmetric ambipolar transfer characteristic, each carrier-dominated flank spans about $\Delta V_b/2$ and its activation barrier changes by about $\Delta E_{\mathrm{SW}}/2$. The thermionic current on either flank therefore changes by approximately $\Delta E_{\mathrm{SW}}/[2\ln(10)k_BT]$ decades, giving the flank-averaged estimate
\begin{equation}
\langle SS\rangle\simeq\ln(10)\,\frac{k_BT}{q}\,\frac{e\,\Delta V_b}{\Delta E_{\mathrm{SW}}},
\label{eq:app-avgss}
\end{equation}
which is 37.7~mV/dec at 300~K and 5.8~mV/dec at 100~K for the $L_b=20$~nm NC-BCFET ($\Delta E_{\mathrm{SW}}/e\Delta V_b=1.58$ and 3.40), but $\approx1.4$~V/dec for the $L_b=20$~nm BCFET ($\Delta E_{\mathrm{SW}}/e\Delta V_b=0.042$). Within this symmetric-flank approximation, $\Delta E_{\mathrm{SW}}/e\Delta V_b>1$ corresponds to a flank-averaged swing below the thermionic value. These window averages are distinct from the tangent-extracted minimum subthreshold swing.

\bibliography{references}

\end{document}


\title{Supporting Information: Sub-Thermionic Switching in a Negative-Capacitance Bandgap- Change Field-Effect Transistor}

\author{Rahul B. Awale}
\affiliation{Department of Electrical Engineering, Indian Institute of Technology Bombay, Powai, Mumbai-400076, India}
\affiliation{Department of Materials Science and Engineering, Monash University, Clayton, Victoria 3800, Australia}

\author{Mike Klymenko}
\affiliation{Department of Materials Science and Engineering, Monash University, Clayton, Victoria 3800, Australia}

\author{Yuefeng~Yin}
\affiliation{Department of Materials Science and Engineering, Monash University, Clayton, Victoria 3800, Australia}

\author{Nikhil V.~Medhekar}
\affiliation{Department of Materials Science and Engineering, Monash University, Clayton, Victoria 3800, Australia}
\affiliation{School of Physics and Astronomy, Monash University, Clayton, Victoria 3800, Australia}

\author{Bhaskaran~Muralidharan}
\affiliation{Department of Electrical Engineering, Indian Institute of Technology Bombay, Powai, Mumbai-400076, India}

\author{Michael~S.~Fuhrer}
\email{Corresponding author: michael.fuhrer@monash.edu}
\affiliation{School of Physics and Astronomy, Monash University, Clayton, Victoria 3800, Australia}
\affiliation{Department of Materials Science and Engineering, Monash University, Clayton, Victoria 3800, Australia}

\date{\today}
\maketitle

This Supporting Information details the device structure, electrostatics, ferroelectric negative-capacitance (NC) formulation, self-consistent iteration framework, density-of-states and transverse-mode-density calculations, channel-potential and drain-current evaluation, subthreshold-swing extraction, and ferroelectric-thickness optimization underlying the bilayer graphene (BLG) NC-BCFET results reported in the main text.

\section*{1. Device Structure and Modeling Framework}

The dual-gated BLG NC-BCFET architecture analyzed in the main text contains an AB-stacked BLG channel sandwiched between a top HfO$_2$ dielectric ($\varepsilon_t \approx 5.98$, thickness $L_t = 10$~nm) and a bottom wurtzite Al$_{1-x}$Sc$_x$N ferroelectric layer at $x = 0.45$ (background dielectric $\varepsilon_b = 3.9$, critical concentration $x_c = 0.6115$).\cite{Gu2024} For benchmarking, we simulate a reference dual-oxide BLG BCFET in which the bottom gate is replaced by SiO$_2$. Al$_{0.55}$Sc$_{0.45}$N is chosen for its CMOS compatibility, large remnant polarization ($\sim$100~$\mu$C/cm$^2$), III--V process integration, and sixth-order Landau--Devonshire (L-D) parameters.\cite{Gu2024}

The ferroelectric is treated within L-D theory in its quasistatic (equilibrium) limit.\cite{Salahuddin2008,Wong2019,Khan2015} The Gibbs free-energy density in the polarization order parameter $P$ is
\begin{equation}
F = \alpha P^{2} + \beta P^{4} + \gamma P^{6} - E_{\mathrm{ext}} P, \tag{1.1}
\end{equation}
where $E_{\mathrm{ext}} = 2\alpha P + 4\beta P^{3} + 6\gamma P^{5}$. Effective Landau coefficients implemented in the series-capacitor model ($V_{\mathrm{FE}} = E_{\mathrm{ext}} t_{\mathrm{FE}}$, $Q \approx P$) are listed in Table~S1. With $\alpha_{\mathrm{FE}} < 0$ and $\beta_{\mathrm{FE}} > 0$, Al$_{0.55}$Sc$_{0.45}$N is a type-I ferroelectric.\cite{Alam2019} The negative-curvature region of $F(P)$ yields a differential capacitance $C_{\mathrm{FE}} = (d^{2}F/dP^{2})^{-1} < 0$;\cite{Salahuddin2008,Wong2019,Cao2020} with $C_{\mathrm{FE}}$ placed in series with the channel and oxide capacitances, the NC regime is stabilized provided the stack-stability margin $M = 1/C_{\mathrm{ox}} + 1/C_{\mathrm{BLG}} + 1/C_{\mathrm{FE}}$ remains positive.\cite{Wong2019,Cao2020,Dong2017}

\begin{table}[H]
\centering
\caption{Effective Landau coefficients for Al$_{0.55}$Sc$_{0.45}$N used in the series-capacitor model.\cite{Gu2024}}
\begin{tabular}{lcc}
\hline
Parameter & Value & Unit \\
\hline
$\alpha_{\mathrm{FE}}$ & $-4.44\times10^{8}$ & $\mathrm{m/F}$ \\
$\beta_{\mathrm{FE}}$ & $3.73\times10^{7}$ & $\mathrm{m^{5}/FC^{2}}$ \\
$\gamma_{\mathrm{FE}}$ & $1.55\times10^{8}$ & $\mathrm{m^{9}/FC^{4}}$ \\
\hline
\end{tabular}
\end{table}

The top gate is fixed at $V_t = +5$~V (large-signal model) to modulate the interlayer asymmetry $\Delta$ and the transport bandgap, while $V_b$ is swept as the switching electrode; this bias regime accesses the full nonlinearity of the ferroelectric polarization required to probe the complete NC operating range. Gauss's law at the dielectric--graphene interfaces [Figure~1(a) of the main text] gives the layer charge densities.\cite{Rahman2003,McCann2013} Throughout the electrostatics we use the interlayer \emph{potential} difference $\Delta_{\phi} \equiv \Delta/e$, in volts, so that it may be combined with the gate voltages $V_{b}$ and $V_{t}$; $\Delta$ itself remains the on-site \emph{energy} asymmetry, in eV, wherever it enters the Hamiltonian, the density of states or the band gap. The layer charge densities are
\begin{equation}
n_{1} = \frac{\varepsilon_{0}\varepsilon_{b}E_{\mathrm{geo}} + P_{f}}{e} + n_{b0}, \tag{1.2}
\end{equation}
\begin{equation}
n_{2} = \frac{\varepsilon_{0}\varepsilon_{t}}{eL_{t}}\left(V_{t} - \frac{\Delta_{\phi}}{2}\right) + n_{t0}, \tag{1.3}
\end{equation}

where $E_{\mathrm{geo}} = (V_{b} + \Delta_{\phi}/2)/L_{b}$, $P_{f}$ is the ferroelectric polarization from Eq.~(1.1), and $n_{t0}, n_{b0}$ are background doping densities (set to zero for intrinsic BLG). At the FE/BLG interface, displacement-field continuity reads $D = \varepsilon_{0}\varepsilon_{b}E + P_{f}$;\cite{Kim2016,Garg2020} the background dielectric constant $\varepsilon_{b}$ regularizes the peak values of $P_{f}$ and sets the strength of the FE--channel electrostatic coupling.

The low-energy electronic structure of BLG arises from the non-dimer sites $A_{1}$ and $B_{2}$.\cite{McCann2013,McCann2006,CastroNeto2009} A perpendicular field applies an on-site offset $\pm\Delta/2$ between layers, lifting the K-point degeneracy and opening a gap that scales as $E_{\mathrm{gap}} \approx \Delta\gamma_{1}/\sqrt{\Delta^{2}+\gamma_{1}^{2}}$ and saturates at $\gamma_{1}$ for $\Delta \gg \gamma_{1}$.\cite{McCann2013,McCann2006} Gate-tunable gaps up to $\sim$250~meV\cite{Zhang2009} and transport bandgaps up to $\sim$130~meV at room temperature\cite{Xia2010,Yan2010} have been demonstrated experimentally in dual-gated BLG FETs.

A single-particle tight-binding (TB) description overestimates the gap because it captures only interlayer screening; the intralayer charge redistribution observed experimentally~\cite{Yan2010,Mak2009,Zhang2009} is absent. The first calculations of this screening effect on the bandgap were reported by Min \textit{et al.}\cite{Min2007} and Zhang \textit{et al.}\cite{Zhang2008IR}, with the direct experimental observation of Zhang \textit{et al.}\cite{Zhang2009} confirming agreement with the screened theory prior to the Corbino-geometry transport measurements of Yan and Fuhrer.\cite{Yan2010} We therefore adopt the \textit{ab initio} GW parametrization,\cite{Gava2009} in which the field-to-gap proportionality $\alpha_{\mathrm{GW}}(n)$ is approximately three times smaller than the bare TB value $\alpha_{\mathrm{bare}} = 30.3\times10^{-12}~\mathrm{cm^{2}\,meV}$. Ballistic transport is treated within the top-of-the-barrier (ToB) formalism.\cite{Rahman2003,Lundstrom2017} The source and drain reservoirs are held at electrochemical potentials $\mu_{1}$ and $\mu_{2} = \mu_{1} - qV_{\mathrm{DS}}$. In contrast to a conventional MOSFET, where switching is driven by barrier-height modulation, here the bandgap itself is the switching variable, tuned by the perpendicular field generated by the dual gates.\cite{McCann2013,McCann2006,CastroNeto2009,Zhang2009,Ohta2006,Castro2007} The drain current follows the L-B formula\cite{Rahman2003,Lundstrom2017}
\begin{equation}
I_{\mathrm{DS}} = \frac{4q}{h}\int M(E)\,T(E)\,[f_{S}(E) - f_{D}(E)]\,dE, \tag{1.4}
\end{equation}
where $T(E) = 1$ in the ballistic infinite-channel limit, $M(E)$ is the number of conducting modes derived from the channel density of states $D(E)$, and the prefactor of 4 accounts for spin and valley degeneracy. The full self-consistent loop couples Eqs.~(1.1)--(1.4) with the GW-corrected $\alpha_{\mathrm{GW}}(n)$ until $\Delta$, $P_{f}$, and $E_{F}$ converge; the closure condition that fixes $E_{F}$ is derived in \S1.1 below.

Figure~S1 shows the cross-sectional schematics of the two device architectures studied in this work. Device~1 (BLG BCFET or oxide device) employs a symmetric double-gate configuration with conventional oxide dielectrics on both sides of the BLG channel. Device~2 (BLG NC-BCFET or NC device) replaces the bottom oxide with a ferroelectric Al$_{0.55}$Sc$_{0.45}$N layer, enabling negative-capacitance operation. 

\begin{figure}[H]
    \centering
    \panel{0.46\linewidth}{a}{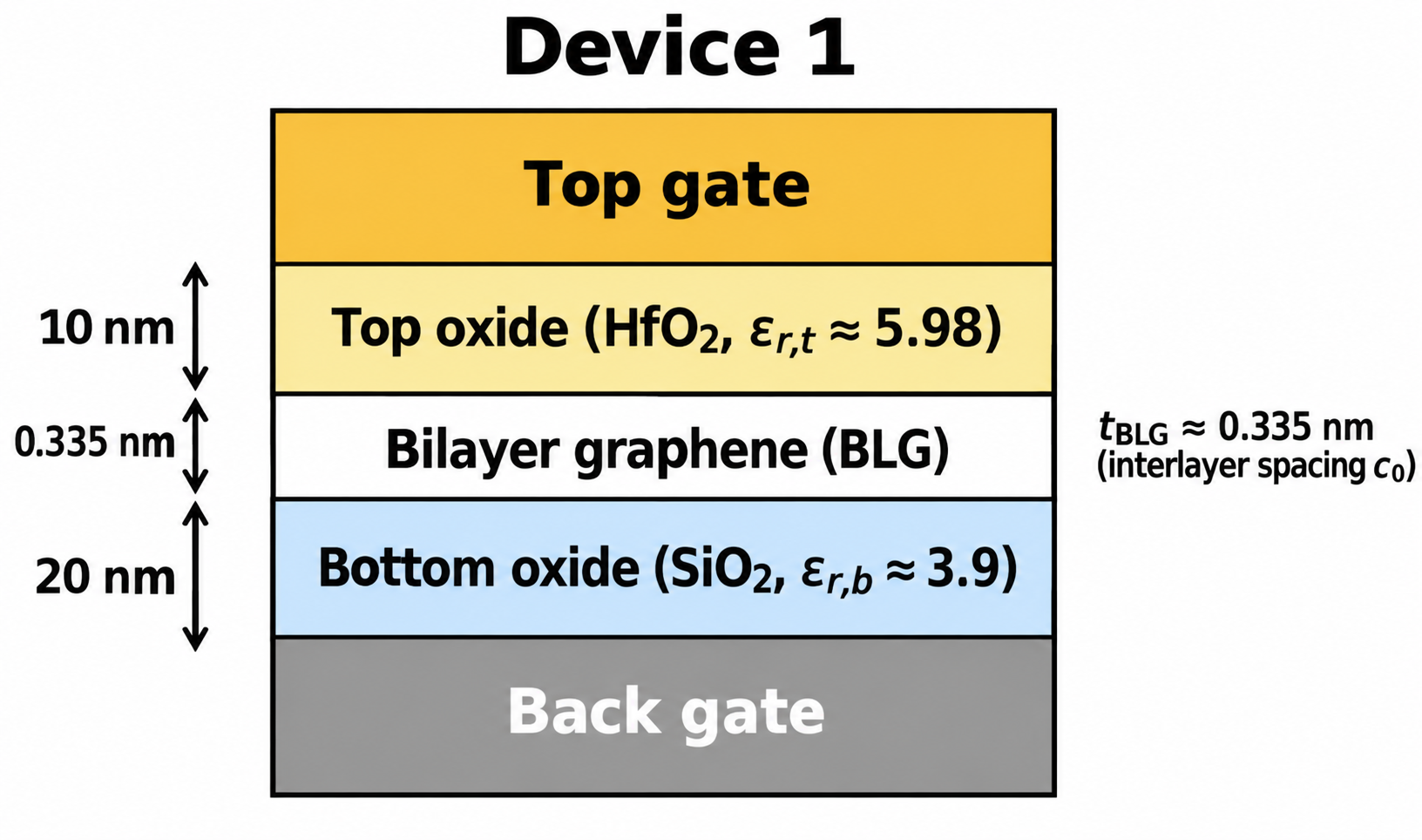}
    \hfill
    \panel{0.46\linewidth}{b}{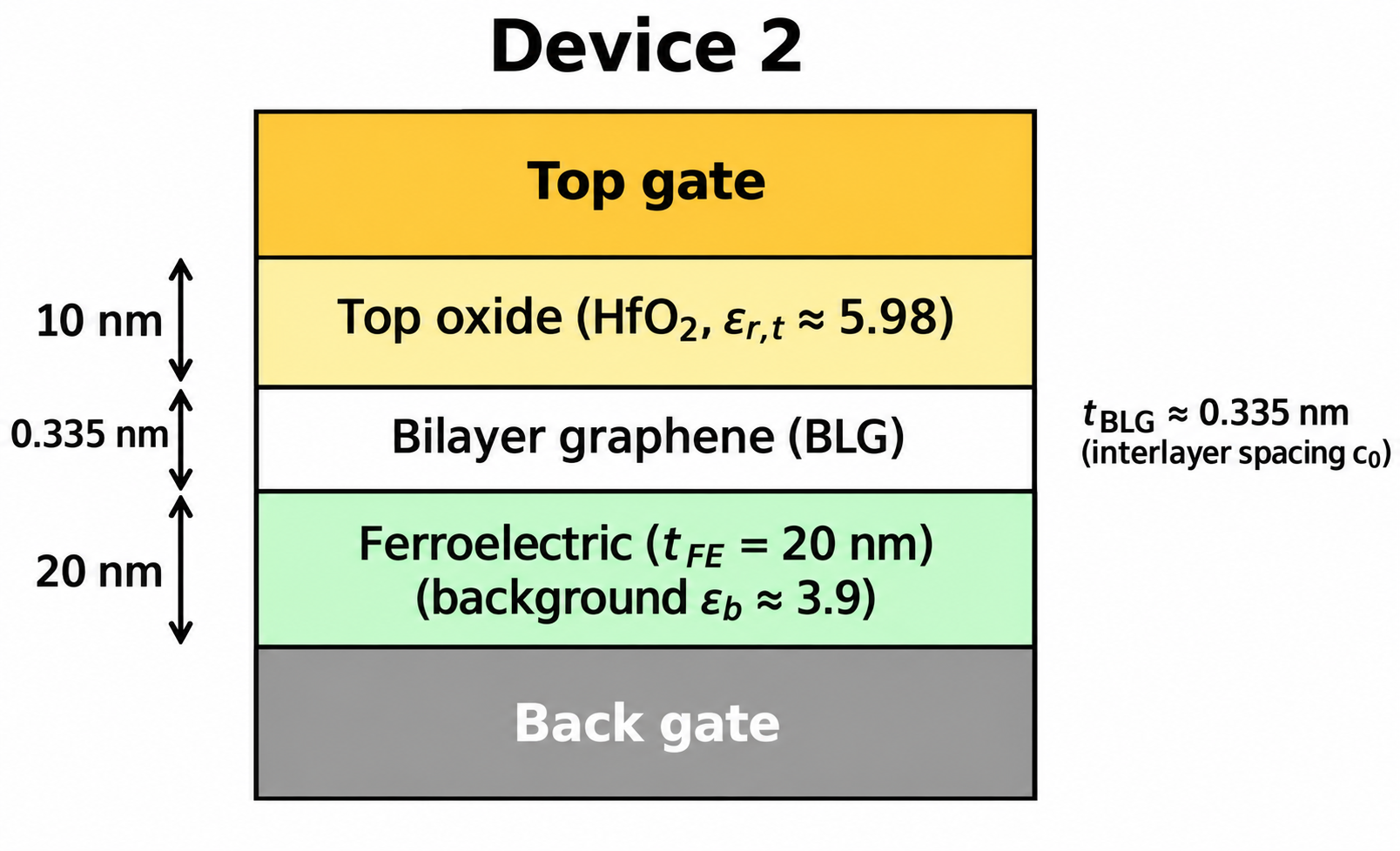}
    \caption{Cross-sectional schematics of the two device architectures. (a) Device~1 (BLG BCFET): double-gated BLG with HfO$_2$ top oxide (10~nm, $\varepsilon_{r,t} = 5.98$) and SiO$_2$ bottom oxide (20~nm, $\varepsilon_{r,b} = 3.9$). (b) Device~2 (BLG NC-BCFET): asymmetric double-gated BLG with HfO$_2$ top oxide and ferroelectric Al$_{0.55}$Sc$_{0.45}$N bottom gate ($t_{\mathrm{FE}} = 20$~nm, background $\varepsilon_b = 3.9$). The BLG interlayer spacing is $c_0 = 0.335$~nm in both devices.}
    \label{fig:S1}
\end{figure}

The equivalent-circuit representation of the BLG NC-BCFET is shown in Figure~S2. This compact model captures the essential capacitive coupling between the top gate ($V_{\mathrm{TG}}$), bottom gate ($V_{\mathrm{BG}}$) with ferroelectric layer, source ($V_S$), drain ($V_{\mathrm{DS}}$), and the BLG channel. The top-gate capacitance $C_{\mathrm{TG}}$ and the bottom-gate capacitance $C_{\mathrm{BG}}$ (which includes the ferroelectric contribution) together control the interlayer asymmetry and the channel potential.

\begin{figure}[H]
    \centering
    \includegraphics[width=0.70\linewidth]{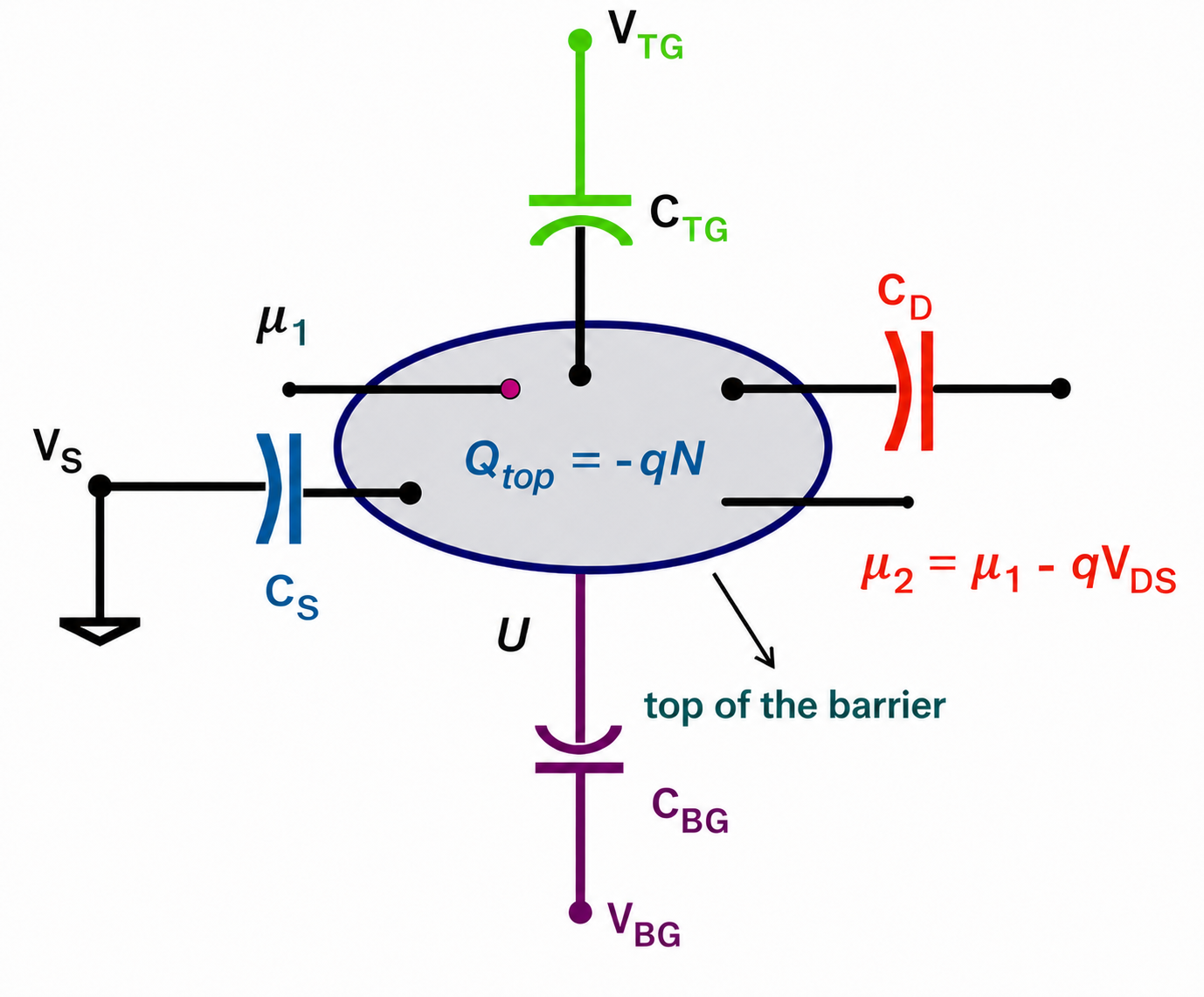}
    \caption{Compact equivalent circuit of the BLG NC-BCFET showing the gate-voltage connections ($V_{\mathrm{TG}}$, $V_{\mathrm{BG}}$), source/drain terminals ($V_S$, $V_{\mathrm{DS}}$), and the BLG channel represented as a capacitive element coupled to both gates. The top-gate capacitance $C_{\mathrm{TG}}$ and bottom-gate capacitance (including ferroelectric) $C_{\mathrm{BG}}$ control the interlayer asymmetry $\Delta$. $U$ is the top-of-the-barrier potential with charge density $Q_{top}$.}
    \label{fig:S2}
\end{figure}

\subsection*{1.1 Channel Potential and the Channel Band Shift}

The gate voltage drops in Eqs.~(1.2)--(1.3) are referenced to charge neutrality, and the common mode is carried entirely by the top-of-the-barrier (ToB) channel potential $U$, which enters the calculation only through the Fermi arguments. The self-consistent cycle is $P_{f} \rightarrow n \rightarrow \Delta \rightarrow U \rightarrow n$, repeated at each bias until $P_{f}$, $n$, $\Delta$ and $U$ have all converged.

At every bias point the Laplacian potential $U_{L}$ of the channel node and the charging energy $U_{0}$ are
\begin{equation}
\phi_{L} = \frac{\bar{C}_{t}\left(V_{t} - \Delta_{\phi}/2\right) + C_{\mathrm{FE}}\left(V_{b} + \Delta_{\phi}/2\right)}{C_{\Sigma}},
\qquad U_{L} = -e\phi_{L}, \qquad U_{0} = \frac{e^{2}}{C_{\Sigma}}, \tag{1.5}
\end{equation}
where $\bar{C}_{t} = \varepsilon_{0}\varepsilon_{t}/L_{t} = 5.29\times10^{-3}~\mathrm{F/m^{2}}$ is the top-oxide capacitance and $C_{\mathrm{FE}}$ is the differential ferroelectric capacitance of Eq.~(3.5), so that the negative curvature of the ferroelectric enters the Laplace drive directly. For the reference BLG BCFET the same expression is used with $C_{\mathrm{FE}}$ replaced by the bottom-oxide capacitance $\bar{C}_{b}$. The charging capacitance is kept geometric in both devices, $C_{\Sigma} = \bar{C}_{t} + \bar{C}_{b} = 7.02\times10^{-3}~\mathrm{F/m^{2}}$ with $\bar{C}_{b} = \varepsilon_{0}\varepsilon_{b}/t_{\mathrm{FE}} = 1.73\times10^{-3}~\mathrm{F/m^{2}}$ (the ferroelectric background permittivity is used for the BLG NC-BCFET), giving $U_{0} = 2.28\times10^{-17}~\mathrm{eV\,m^{2}}$, i.e.\ 228~meV per $10^{16}~\mathrm{m^{-2}}$; this keeps $U_{0} > 0$ so that the $U$-solve stays single-valued while the negative capacitance is retained in $\phi_{L}$. The channel potential then follows from the ToB charge balance $U = U_{L} + U_{0}(n - n_{0})$ of Eq.~(6.4), in which the quantum charge is evaluated with $U$ inside the Fermi function,
\begin{equation}
n = \int D_{0}(E;\Delta)\, f\!\left(E + U - \mu_{0}\right) dE - n_{\mathrm{ref}},
\qquad E_{F,\mathrm{local}} = \mu_{0} - U, \tag{1.6}
\end{equation}
where $\mu_{0} = 0$ is the grounded-source reference. This filling is the point at which the net sheet charge of Eq.~(1.6) vanishes and is what $E_{\mathrm{CNP}}$ denotes throughout; it coincides with the trace-zero on-site mean of $H_{0}(\Delta)$, and because $\gamma_{4}$ renders the spectrum electron--hole asymmetric it sits within $\sim$10~meV of the strict $n_{e} = n_{h}$ level. The offset is common to both band edges and therefore cancels in the transport gap and in every energy difference quoted here. It is distinct from the band-edge midpoint $(E_{c0}+E_{v0})/2$, which is not a neutrality point. The source electrochemical potential is the fixed reference, $\mu_{0} = 0$, and the channel bands shift rigidly by $U$, so that $n = \int D_{0}(E-U;\Delta)\,f(E-\mu_{0})\,dE - n_{\mathrm{ref}}$; this is the standard top-of-the-barrier form,\cite{Datta2005} in which an increase in $U$ raises the channel levels and depletes the channel. Equation~(1.6) therefore identifies the channel band shift\cite{Zhi2020} with the self-consistent ToB channel potential:
\begin{equation}
\Delta E = E_{F} - E_{\mathrm{CNP}} = -U. \tag{1.7}
\end{equation}
This is the feedback path used throughout this work. The scalar $n$ of Eq.~(1.6) is returned to the L-D solver, to the GW coefficient $\alpha_{\mathrm{GW}}(n)$ and hence to $\Delta$, so that $U$ modifies the polarization, the screening and the transport band gap.

\subsection*{1.2 GW Screening Parametrization}

The interlayer asymmetry is obtained from the layer densities with the GW-screened coupling of Gava \textit{et al.},\cite{Gava2009}
\begin{equation}
\Delta = \alpha_{\mathrm{GW}}(n)\,(n_{2} - n_{1}), \qquad
\alpha_{\mathrm{GW}}(n) = \sum_{i=1}^{3} \frac{A_{i}}{1 + \left[(n - B_{i})/\Gamma_{i}\right]^{2}} + C, \tag{1.8}
\end{equation}
where $n$ is the mobile sheet density of Eq.~(1.6), taken from the previous outer iteration and expressed in units of $10^{12}~\mathrm{cm^{-2}}$. The parameters are listed below; $A_{i}$ and $C$ are in units of $10^{-12}~\mathrm{cm^{2}\,meV} = 10^{-19}~\mathrm{eV\,m^{2}}$, the unit in which $\alpha_{\mathrm{GW}}$ is reported in Table~S2.

\begin{center}
\begin{tabular}{cccc}
\hline
$i$ & $A_{i}$ ($10^{-12}~\mathrm{cm^{2}\,meV}$) & $B_{i}$ ($10^{12}~\mathrm{cm^{-2}}$) & $\Gamma_{i}$ ($10^{12}~\mathrm{cm^{-2}}$) \\
\hline
1 & 0.896 & $-26.888$ & 21.756 \\
2 & 3.905 & 1.623 & 21.946 \\
3 & $-1.654$ & $-0.092$ & 5.534 \\
$C$ & 5.848 & -- & -- \\
\hline
\end{tabular}
\end{center}

\section*{2. Electrostatics and Charge Coupling in Devices}

The device has an AB-stacked BLG channel between a top gate (HfO$_2$, $\varepsilon_{r,t} = 5.98$, thickness $L_t = 10$~nm)\cite{Chen2025} and a bottom gate. The dielectric in the BLG BCFET is SiO$_2$ ($\varepsilon_{r,b} = 3.9$, $L_b = 20$~nm); the dielectric in the BLG NC-BCFET is a ferroelectric Al$_{0.55}$Sc$_{0.45}$N layer with a background dielectric constant $\varepsilon_b = 3.9$ and a thickness $t_{\mathrm{FE}}$ between 8 and 20~nm. The voltage on the upper gate, $V_t$, stays at $+5$~V, while the voltage on the lower gate, $V_b$, changes. Using Gauss's law at the dielectric--graphene interfaces,\cite{Rahman2003,McCann2013} the charge densities on the lower (layer~1) and upper (layer~2) graphene layers are, with $\Delta_{\phi} = \Delta/e$ as defined in \S1:

\noindent BLG BCFET:
\begin{gather}
n_{1} = \frac{\varepsilon_{0}\varepsilon_{r,b}}{eL_{b}}\left(V_{b} + \frac{\Delta_{\phi}}{2}\right) + n_{b0}, \tag{2.1}\\[2pt]
n_{2} = \frac{\varepsilon_{0}\varepsilon_{r,t}}{eL_{t}}\left(V_{t} - \frac{\Delta_{\phi}}{2}\right) + n_{t0}. \tag{2.2}
\end{gather}
\noindent BLG NC-BCFET:
\begin{equation}
n_{1} = \frac{\varepsilon_{0}\varepsilon_{b}E_{\mathrm{geo}} + P_{f}}{e} + n_{b0}, \tag{2.3}
\end{equation}
where $E_{\mathrm{geo}} = (V_{b} + \Delta_{\phi}/2)/t_{\mathrm{FE}}$ is the geometric electric field in the lower dielectric, and $P_{f}$ is the ferroelectric polarization that follows the L-D equation (Section~3). For intrinsic BLG, $n_{b0} = n_{t0} = 0$. The fundamental distinction between the two devices is depicted in Eq.~(2.3): the carrier density in the bottom layer includes the ferroelectric polarization $P_{f}$. Note that none of Eqs.~(2.1)--(2.3) contains the channel potential. The channel band shift that accompanies gate-induced charging of the bilayer, $\Delta E$,\cite{Zhi2020} is carried in full by the ToB channel potential through $\Delta E = -U$, as derived in \S1.1.

\subsection*{2.1 Profile of Electrostatic Potential}

As Laplace's equation holds in each charge-free dielectric region, the electrostatic potential $\phi(x)$ across the vertical device cross-section is linear. The potential in each region is set by the bottom-gate position, which is $x = -L_{b}$, the BLG layers, which are at $x = \pm c_{0}/2$, and the top gate, which is at $x = +L_{t}$.

\noindent Region~I (bottom dielectric, $x \in [-L_{b}, -c_{0}/2]$):
\begin{equation}
\phi^{\mathrm{I}}(x) = V_{b} - \frac{V_{b} + \Delta_{\phi}/2}{L_{b} - c_{0}/2}\,(x + L_{b}). \tag{2.4}
\end{equation}
\noindent Region~II (BLG interlayer, $x \in [-c_{0}/2, +c_{0}/2]$):
\begin{equation}
\phi^{\mathrm{II}}(x) = \frac{\Delta_{\phi}}{c_{0}}\,x. \tag{2.5}
\end{equation}
\noindent Region~III (top dielectric, $x \in [+c_{0}/2, +L_{t}]$):
\begin{equation}
\phi^{\mathrm{III}}(x) = \frac{\Delta_{\phi}}{2} + \frac{V_{t} - \Delta_{\phi}/2}{L_{t} - c_{0}/2}\left(x - \frac{c_{0}}{2}\right). \tag{2.6}
\end{equation}
The two dielectric regions span $L_{b} - c_{0}/2$ and $L_{t} - c_{0}/2$ rather than $L_{b}$ and $L_{t}$, because the graphene sheets sit at $x = \pm c_{0}/2$ and not at the origin; Eqs.~(2.4) and~(2.6) use these exact widths, so that each profile reproduces both of its boundary values. The capacitances in Eqs.~(1.5) and~(2.1)--(2.3) instead use $L_{b}$, $L_{t}$ and $t_{\mathrm{FE}}$, i.e.\ the separation between each gate electrode and the \emph{centre} of the channel, which is the distance that defines the insulator capacitance $C_{\mathrm{ins}} = \varepsilon/d$ of the top-of-the-barrier model.\cite{Datta2005} The two sets of lengths therefore differ by design: the profile equations use the physical extent of each dielectric, whereas the capacitances use the gate-to-channel-centre separation. Figure~2 of the main text plots the assembled profile $\phi_{\mathrm{es}}(x)$---that is, Eqs.~(2.4)--(2.6) written as a single function of position---in volts; the axes follow Figure~9(b) of Ref.~\citenum{fuhrer2021nctqfet}, which shows the corresponding potential energy $-e\phi_{\mathrm{es}}(x)$. Note that $\phi_{\mathrm{es}}(x)$ is the Laplace potential across the stack and is distinct from the scalar top-of-the-barrier channel potential $U$ of \S1.1: Eq.~(2.5) gives $\phi_{\mathrm{es}}(0) = 0$ at every bias, whereas $U \neq 0$ carries the common-mode band shift. When NC is active, the sign of $\Delta$ changes, which flips the potential profile in the BLG region (Eq.~2.5). This flip is the physical signature of the NC cascade and can be seen directly in Figure~2 of the main text. The switching window (SW) is the range of $V_{b}$ values over which the Fermi level crosses the entire transport band gap,\cite{Ohta2006,Castro2007} with the two SW ON states on either side. The measured SW widths are 4.79~V (BLG BCFET, 300~K), 0.128~V (BLG NC-BCFET, 300~K), and 0.059~V (BLG NC-BCFET, 100~K); the SW edges lie at $V_{b} = -17.51$ and $-12.72$~V for the BLG BCFET, at $+0.27$ and $+0.40$~V for the BLG NC-BCFET at 300~K, and at $+0.31$ and $+0.37$~V at 100~K.

\section*{3. Formulation of Ferroelectric Negative Capacitance}

\subsection*{3.1 Landau--Devonshire Free Energy and Polarization}

The sixth-order L-D thermodynamic energy density\cite{Gu2024,Salahuddin2008,Wong2019} describes how the Al$_{0.55}$Sc$_{0.45}$N bottom gate behaves as a ferroelectric:
\begin{equation}
F = \alpha P^{2} + \beta P^{4} + \gamma P^{6} - E_{\mathrm{ext}} P, \tag{3.1}
\end{equation}
where $\alpha$, $\beta$, and $\gamma$ are the Landau coefficients that depend on the composition of wurtzite Al$_{1-x}$Sc$_x$N. At Sc concentration $x = 0.45$ with critical concentration $x_c = 0.6115$, the effective coefficients are
\begin{gather}
\alpha_{\mathrm{FE}} = \frac{\alpha_{0}(x - x_c)}{2} \approx -4.44\times10^{8}~\mathrm{m/F}, \tag{3.2a}\\[2pt]
\beta_{\mathrm{FE}} = \frac{\gamma_{\mathrm{paper}}}{4} \approx 3.73\times10^{7}~\mathrm{m^{5}\,F^{-1}\,C^{-2}}, \tag{3.2b}\\[2pt]
\gamma_{\mathrm{FE}} = \frac{\omega_{\mathrm{paper}}}{6} \approx 1.55\times10^{8}~\mathrm{m^{9}\,F^{-1}\,C^{-4}}, \tag{3.2c}
\end{gather}
where $\alpha_{0} = 5.504\times10^{9}~\mathrm{m/F}$, $\gamma_{\mathrm{paper}} = 0.149\times10^{9}~\mathrm{m^{5}\,F^{-1}\,C^{-2}}$, and $\omega_{\mathrm{paper}} = 0.929\times10^{9}~\mathrm{m^{9}\,F^{-1}\,C^{-4}}$. The material is a type-I ferroelectric because $\alpha_{\mathrm{FE}} < 0$ and $\beta_{\mathrm{FE}} > 0$. This means it has a polarization--electric-field curve with a negative-slope region that corresponds to the NC regime. At thermodynamic equilibrium ($dP/dt = 0$), minimizing the free energy~\cite{Salahuddin2008,Wong2019} gives the implicit relationship between the external electric field and the equilibrium polarization $P_{f}$:
\begin{equation}
E_{\mathrm{ext}} = 2\alpha_{\mathrm{FE}} P_{f} + 4\beta_{\mathrm{FE}} P_{f}^{3} + 6\gamma_{\mathrm{FE}} P_{f}^{5}. \tag{3.3}
\end{equation}
Setting $E_{\mathrm{ext}}$ equal to the geometric field $E_{\mathrm{geo}} = (V_{b} + \Delta_{\phi}/2)/t_{\mathrm{FE}}$, Eq.~(3.3) becomes a quintic equation in $P_{f}$. Landau symmetry states that only odd powers of $P_{f}$ appear, yielding the polynomial
\begin{equation}
6\gamma_{\mathrm{FE}} P_{f}^{5} + 4\beta_{\mathrm{FE}} P_{f}^{3} + 2\alpha_{\mathrm{FE}} P_{f} - E_{\mathrm{geo}} = 0. \tag{3.4}
\end{equation}
At each bias point, this quintic is solved for all real roots. Each real root is screened with the stack-stability margin of Eq.~(3.6), evaluated with the geometric $C_{\mathrm{BLG,diff}}$ of \S8.4. Below the critical ferroelectric-insulator thickness $t_{c} = 1/(2|\alpha_{\mathrm{FE}}|C_{\mathrm{ox}}) = 212$~nm, the origin can be stabilized in the NC regime; the screened root with the smallest $|P_{f}|$ is selected, and below the coercive field this root lies on the negative-slope branch of the S-curve.\cite{Salahuddin2008,Wong2019} Above $t_{c}$, the origin loses stability and hysteretic operation results according to the critical-thickness criterion of Salahuddin and Datta; the screened root nearest the previous polarization would then be selected.\cite{Salahuddin2008,Alam2019,Khan2015} All devices studied here ($t_{\mathrm{FE}} = 8$--20~nm) use the NC-stabilized selection. 

\subsection*{3.2 The Displacement Field at the FE/BLG Interface}

The total polarization of the ferroelectric layer has two parts: (i) the background lattice polarization $P_{b} = \varepsilon_{0}(\varepsilon_{b} - 1)E$, which describes how the crystal lattice responds to an electric field, and (ii) the spontaneous ferroelectric polarization $P_{f}$, which arises from collective ionic displacement. The total displacement field at the FE/BLG interface is therefore $D = \varepsilon_{0}\varepsilon_{b}E + P_{f}$. The background dielectric constant $\varepsilon_{b} = 3.9$ prevents $P_{f}$ from attaining unrealistically large values and ensures that the BLG channel and the FE gate are properly electrostatically coupled.

\subsection*{3.3 NC Stability and Capacitance Examination}

The ferroelectric differential capacitance per unit area is obtained from the second derivative of the Landau free energy:
\begin{equation}
C_{\mathrm{FE}} = \frac{\varepsilon_{0}\varepsilon_{b} + 1/\kappa}{t_{\mathrm{FE}}}, \tag{3.5}
\end{equation}
where $\kappa = d^{2}F/dP^{2} = 2\alpha_{\mathrm{FE}} + 12\beta_{\mathrm{FE}} P_{f}^{2} + 30\gamma_{\mathrm{FE}} P_{f}^{4}$ is the Landau curvature. Near $P_{f} = 0$, where the double-well free-energy landscape has negative curvature ($\kappa < 0$), the incremental capacitance $C_{\mathrm{FE}}$ becomes negative. This is the origin of the NC effect.\cite{Salahuddin2008,Wong2019} For the quasistatic operating point to stay stabilized against spontaneous polarization switching, the stack-stability margin must stay positive:
\begin{equation}
M = \frac{1}{C_{\mathrm{ox}}} + \frac{1}{C_{\mathrm{BLG}}} + \frac{1}{C_{\mathrm{FE}}} > 0, \tag{3.6}
\end{equation}
where $C_{\mathrm{ox}} = \varepsilon_{0}\varepsilon_{r,t}/L_{t}$ is the top-oxide capacitance and $C_{\mathrm{BLG}}$ is the BLG channel capacitance. There are three types of bias points: Case~A ($C_{\mathrm{FE}} > 0$), which means normal ferroelectric operation without NC; Case~B ($C_{\mathrm{FE}} < 0$ with $M > 0$), which means stabilized NC where sub-60~mV/dec switching is possible; and Case~C ($C_{\mathrm{FE}} < 0$ with $M \le 0$), in which the quasistatic branch is not stabilized and a dynamic treatment would be expected to show hysteretic switching.\cite{Cao2020}

\subsection*{3.4 The NC Cascade and Voltage Amplification}

The NC cascade proceeds as follows: the negative Landau curvature ($d^{2}F/dP^{2} < 0$) makes $C_{\mathrm{FE}} < 0$, which changes the sign of the voltage drop across the ferroelectric layer $V_{\mathrm{FE}}$. This change flips the geometric electric field $E_{\mathrm{geo}}$, which then flips the interlayer asymmetry $\Delta$ through the self-consistent electrostatic loop. As a result, the channel charge-neutrality point rises above the source Fermi level (equivalently, $E_{F,\mathrm{local}}$ drops below the CNP; \S1.1), and the change in surface potential exceeds the change in gate voltage, so the voltage amplification factor $A_V = d\psi_s/dV_g > 1$. The subthreshold swing (SS) under NC operation is
\begin{equation}
SS = \frac{\partial V_{g}}{\partial\psi_{s}}\cdot\frac{\partial\psi_{s}}{\partial(\log_{10}|I_{d}|)} = m\,n, \tag{3.7}
\end{equation}
where $m = \partial V_{g}/\partial\psi_{s} = 1 + C_{s}/C_{\mathrm{ins}}$ is the dimensionless body factor and $n = \partial\psi_{s}/\partial(\log_{10}|I_{d}|)$ is the transport factor, which carries the units of $SS$. For ideal thermionic injection the transport factor takes its Boltzmann value $n = \ln(10)\,k_{B}T/q \approx 59.5$~mV/dec at 300~K, so that $SS = 59.5\,m$~mV/dec. When $C_{\mathrm{ins}} < 0$, the body factor $m < 1$ and $SS$ falls below the Boltzmann limit.\cite{Salahuddin2008,Wong2019}

\subsection*{3.5 Ferroelectric Polarization in Relation to Gate Voltage}

The ferroelectric polarization $P_{f}$, derived from solving the quintic Eq.~(3.4) at each self-consistent bias point, exhibits the polarization dependence on the bottom-gate voltage $V_{b}$ anticipated by Landau theory. The operating point moves along the $P$--$E$ curve as $V_{b}$ is swept. In the NC operating region, where $C_{\mathrm{FE}} < 0$ and the stability margin $M > 0$ (Eq.~3.6), the polarization moves through the thermodynamically unstable branch of the S-curve. This is the negative-slope part of the curve that connects the two local free-energy minima. The BLG channel capacitance $C_{\mathrm{BLG}}$ and the top-oxide capacitance $C_{\mathrm{ox}}$ together stabilize this branch, preventing the system from switching to a stable polarization state on its own~\cite{Salahuddin2008,Wong2019}. In the normal positive-capacitance mode, the ferroelectric operates outside the NC region, and $P_{f}$ lies on one of the two stable branches. The transition from normal to NC operation is smooth as long as $M$ stays positive. The overall polarization at the FE/BLG interface is the sum of the switchable ferroelectric part and the linear background lattice response:
\begin{equation}
P_{\mathrm{total}} = P_{f} + \varepsilon_{0}(\varepsilon_{b} - 1)E_{\mathrm{geo}}. \tag{3.8}
\end{equation}
The Landau equation gives the internal electric field inside the ferroelectric layer as $E_{\mathrm{FE}} = 2\alpha_{\mathrm{FE}} P_{f} + 4\beta_{\mathrm{FE}} P_{f}^{3} + 6\gamma_{\mathrm{FE}} P_{f}^{5}$. The associated voltage drop is $V_{\mathrm{FE}} = E_{\mathrm{FE}}\,t_{\mathrm{FE}}$. When the charge increases in the NC regime, $V_{\mathrm{FE}}$ decreases ($dV_{\mathrm{FE}}/dQ < 0$). This is the negative-differential-capacitance phenomenon.\cite{Salahuddin2008,Khan2015}

\section*{4. Representative Bias-Point Band Structure and Screening Values}

Table~S2 consolidates the converged interlayer asymmetry $\Delta$, transport band gap $E_{\mathrm{gap}}$, and GW screening coefficient $\alpha_{\mathrm{GW}}$ at the OFF, SW~ON$_1$, SW~ON$_2$, and ON bias points reported in the Results and discussion section of the main text. The values are obtained from the self-consistent solver of \S4 combined with the Lorentzian GW parametrization of Eq.~(1.8), and they fix the electronic structure used by the DOS, mode-density, and ToB transport calculations of \S\S5--6.

\begin{table}[H]
\centering
\small
\caption{Representative bias-point band-structure values for the three device configurations of the main text. $\alpha_{\mathrm{GW}}$ is reported in units of $10^{-19}~\mathrm{eV\cdot m^{2}}$ (see \S1.2 for unit conversion). A negative $\Delta$ at the FE ON state ($V_b = -0.70$~V) signals the NC-driven inversion of the interlayer asymmetry. The FE 100~K OFF row is identical to the 300~K OFF row because the large-bias electrostatics are temperature-insensitive (main text, Results and discussion).}
\begin{tabular}{lcccc}
\hline
Device / state & $V_b$ (V) & $E_{\mathrm{gap}}$ (meV) & $\Delta$ (meV) & $\alpha_{\mathrm{GW}}$ ($\times10^{-19}$) \\
\hline
Oxide 300~K --- OFF & $-20.0$ & 253.4 & 333.4 & 8.93 \\
Oxide 300~K --- SW ON$_1$ & $-17.51$ & 237.0 & 298.4 & 8.59 \\
Oxide 300~K --- SW ON$_2$ & $-12.72$ & 214.2 & 256.4 & 8.64 \\
Oxide 300~K --- ON & $-0.70$ & 138.8 & 148.5 & 8.76 \\
FE 300~K --- OFF & $+0.50$ & 261.4 & 352.3 & 9.05 \\
FE 300~K --- SW ON$_1$ & $+0.40$ & 236.4 & 297.2 & 8.58 \\
FE 300~K --- SW ON$_2$ & $+0.27$ & 214.2 & 256.4 & 8.64 \\
FE 300~K --- zero bias & $0.00$ & 156.8 & 171.2 & 8.95 \\
FE 300~K --- ON & $-0.70$ & 55.7 & $-56.3$ & 6.90 \\
FE 100~K --- OFF & $+0.50$ & 261.4 & 352.3 & 9.05 \\
FE 100~K --- SW ON$_1$ & $+0.37$ & 228.0 & 281.1 & 8.44 \\
FE 100~K --- SW ON$_2$ & $+0.31$ & 217.8 & 262.6 & 8.47 \\
FE 100~K --- ON & $-0.70$ & 55.7 & $-56.3$ & 6.90 \\
\hline
\end{tabular}
\end{table}

\section*{5. Density of States and Transverse Mode Density}

\subsection*{5.1 Density of States}

Using Gaussian broadening, the density of states $D(E)$ is obtained from the full four-band eigenvalue spectrum\cite{McCann2013}:
\begin{equation}
D(E) = \frac{4\,\Delta k_{z}\,\Delta k_{y}}{(2\pi)^{2}}\sum_{k}\sum_{\nu}\delta_{\gamma}\big(E - E_{\nu}(k)\big), \tag{5.1}
\end{equation}
where the prefactor $4\,\Delta k_{z}\,\Delta k_{y}/(2\pi)^{2}$ is the $k$-space integration weight including the fourfold spin--valley degeneracy, the sum extends over all $k$-points and all four bands $\nu$, and $\delta_{\gamma}$ is a normalized Gaussian broadening function:
\begin{equation}
\delta_{\gamma}(\varepsilon) = \frac{1}{\gamma\sqrt{2\pi}}\exp\!\left(-\frac{\varepsilon^{2}}{2\gamma^{2}}\right), \tag{5.2}
\end{equation}
with a broadening width $\gamma = 1.5$~meV. 

\subsection*{5.2 Density of Transverse Modes}

The transverse mode density $M(E)$ counts the number of right-moving propagating modes available to carry current at energy $E$.\cite{Rahman2003,Lundstrom2017} It follows from the group velocity of all four bands:
\begin{equation}
v_{z,\nu}(k) = \frac{e}{\hbar}\,\frac{\partial E_{\nu}}{\partial k_{z}}. \tag{5.3}
\end{equation}
The mode density is then computed as a dimensionless sum over right-moving states:
\begin{equation}
M(E) = \sum_{k}\sum_{\nu}\delta_{\gamma}\big(E - E_{\nu}(k)\big)\,\Theta(v_{z,\nu})\,\frac{\hbar\,|v_{z,\nu}|}{e}\,\Delta k_{z}, \tag{5.4}
\end{equation}
where $\Theta$ is the Heaviside step function that selects only states with positive $z$-velocity (moving to the right). Equation~(5.4) sums over the stored $k_{y}$ grid without a $1/2\pi$ weighting, so $M(E)$ is a dimensionless mode \emph{count} for the implicit transverse width $W = 2\pi/\Delta k_{y}$ of that grid, and not a mode density per unit width. Currents derived from it are therefore divided by $W$ before any two devices are compared, because the BLG BCFET and BLG NC-BCFET data sets were stored on $k_{y}$ grids of different spacing and so carry different implicit widths ($W = 1.082~\mu$m and $2.167~\mu$m, respectively; main text, Figure~3). The mode count is computed from all four bands so that it is consistent with the full-band DOS.

\section*{6. CNP-referenced Fermi energy \texorpdfstring{$\Delta E$}{Delta E} and the Drain Current}

\subsection*{6.1 Fermi energy level}

\begin{figure}[H]
    \centering
    \panel{0.32\linewidth}{a}{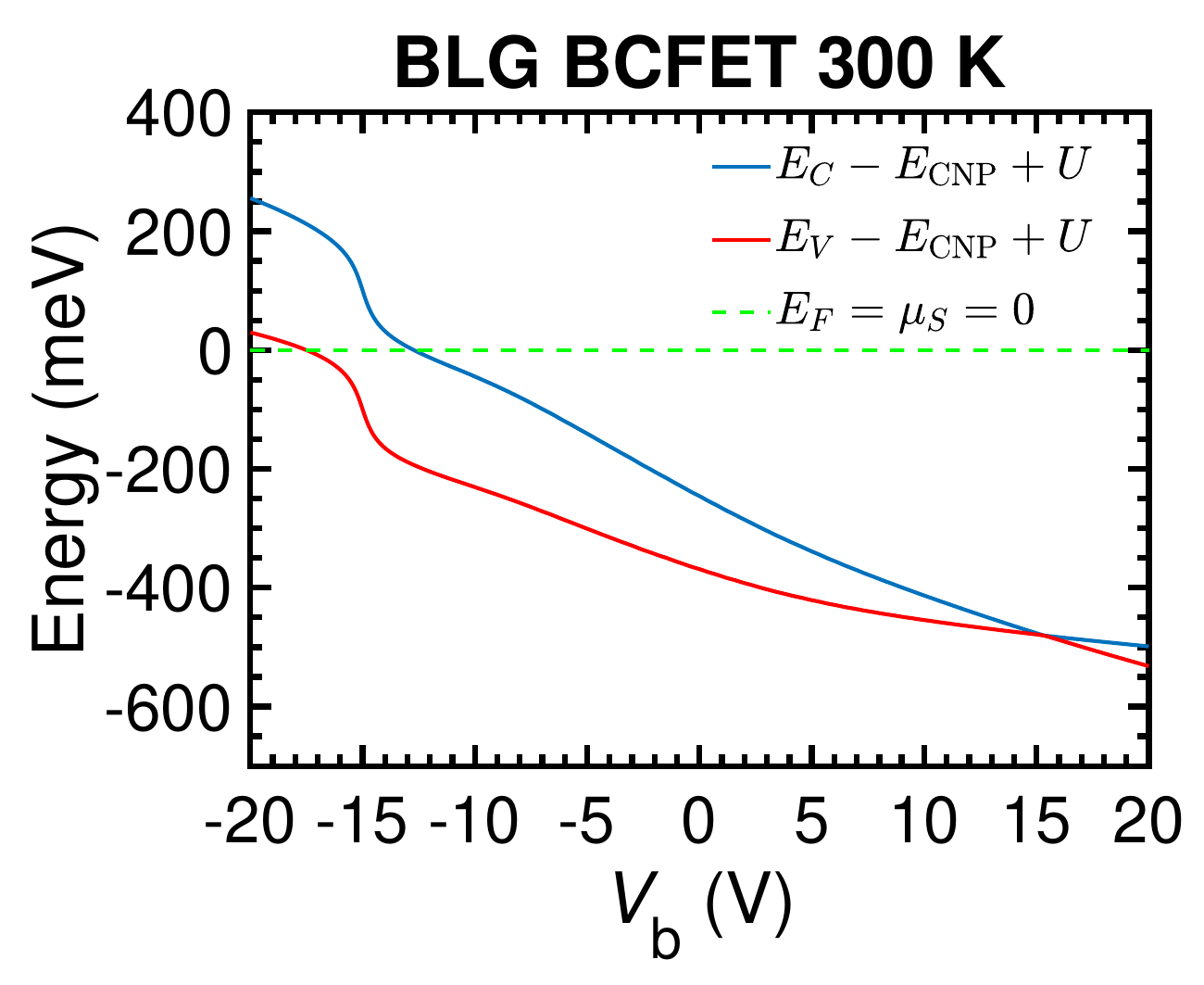}
    \hfill
    \panel{0.32\linewidth}{b}{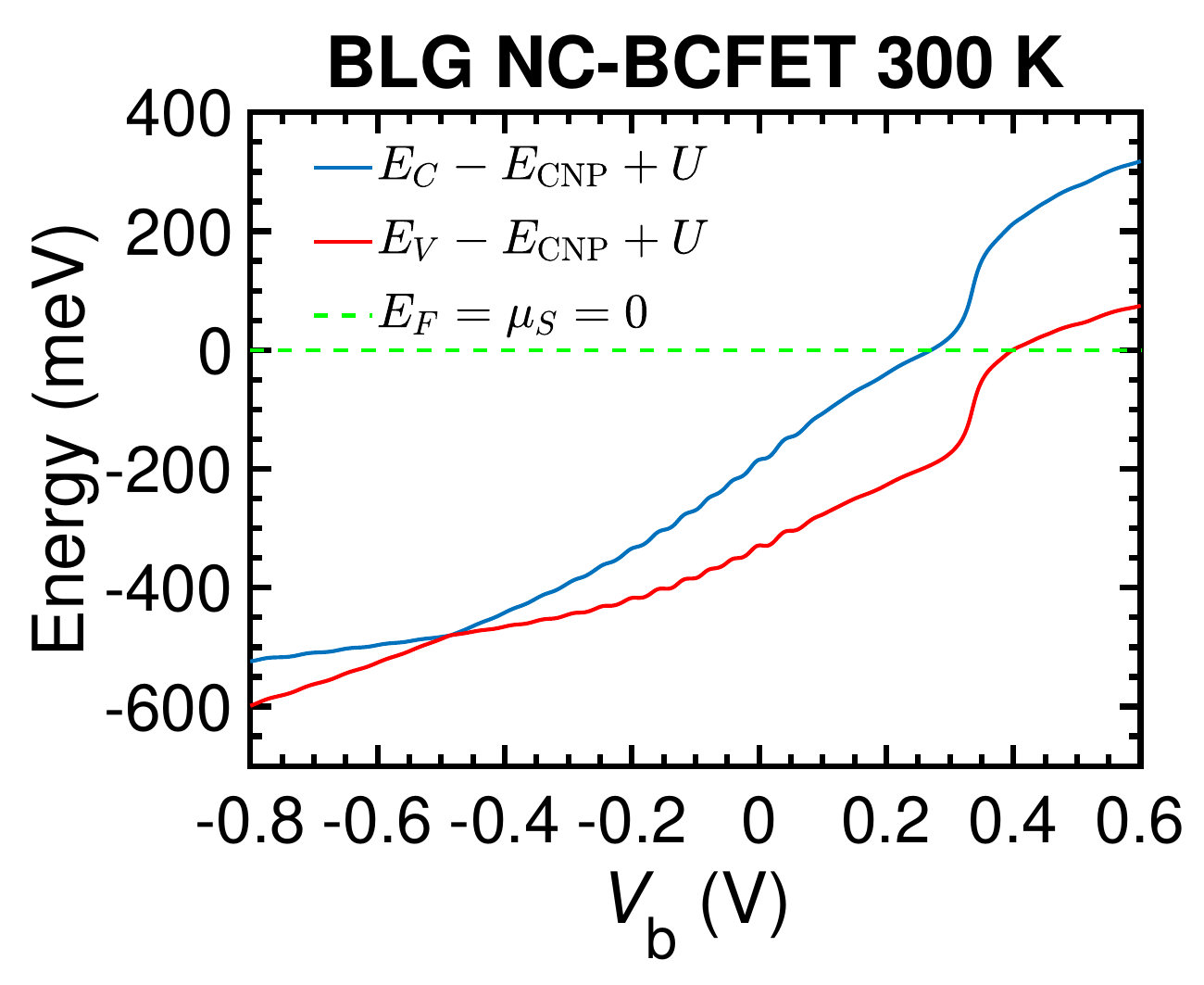}
    \hfill
    \panel{0.32\linewidth}{c}{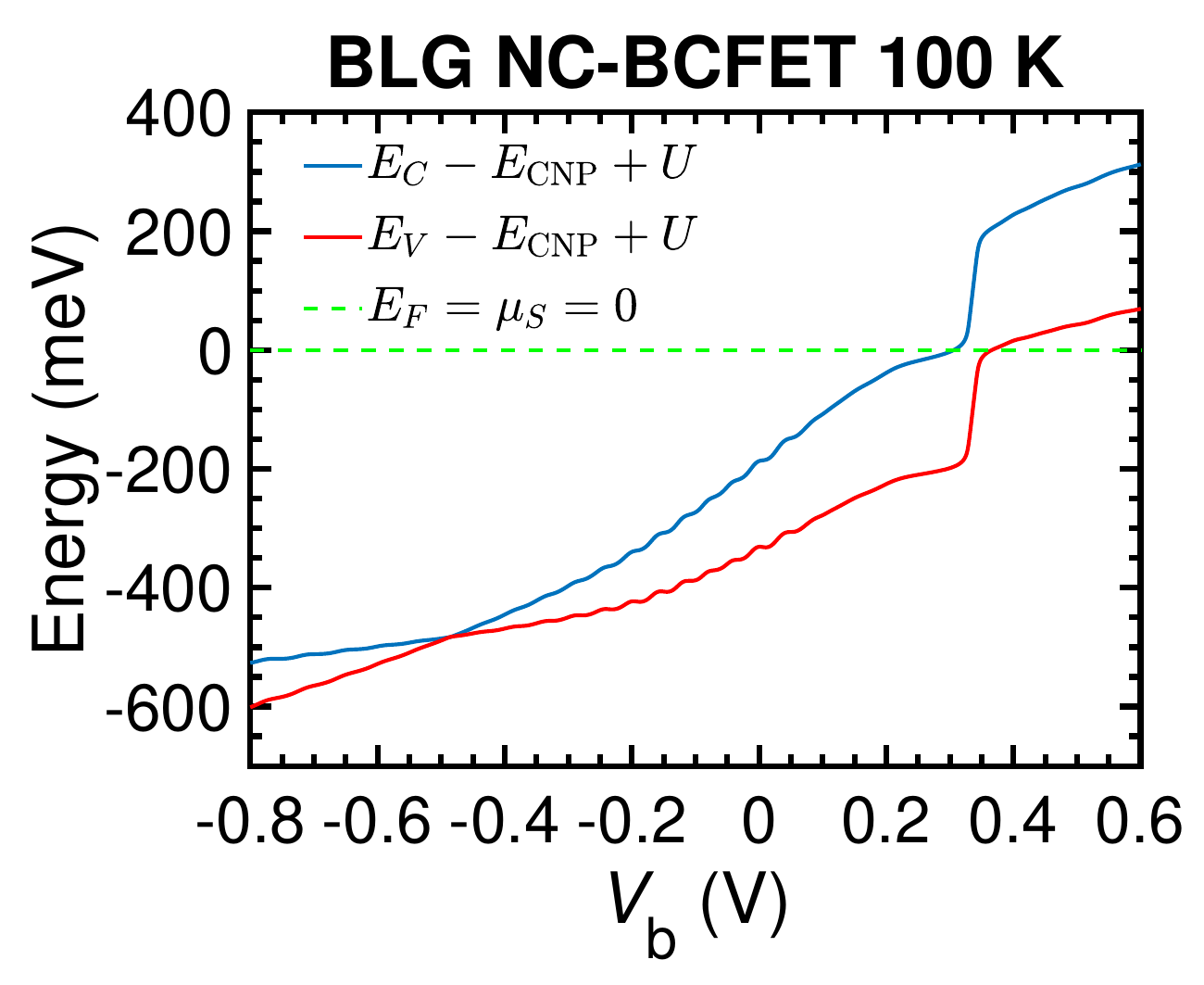}
    \caption{Laboratory-frame view of the data plotted in Figure~2(b,d,f) of the main text. Here the
    grounded source pins the reservoir Fermi level at $E_{F} = \mu_{S} = 0$\cite{Rahman2003} and the channel
    bands carry the common-mode shift, so the absolute band edges are $E_C - E_{\mathrm{CNP}} + U$ and $E_V - E_{\mathrm{CNP}} + U$. 
    (a)~BLG BCFET at 300~K, (b)~BLG NC-BCFET at 300~K, (c)~BLG NC-BCFET at 100~K. This is the same solution as
    the main-text Figure~2, re-referenced; the two differ only by the rigid shift $U$ and must not be
    combined.}
    \label{fig:S3}
\end{figure}

In every calculation reported here the channel-potential feedback is active, and the Fermi level is not solved for: the channel occupation is fixed by the self-consistent top-of-the-barrier potential, so that
\begin{equation}
E_{F,\mathrm{local}} = \mu_{0} - U, \qquad \Delta E = E_{F} - E_{\mathrm{CNP}} = -U, \tag{6.1}
\end{equation}
with $\mu_{0} = 0$ the grounded-source reference of \S1.1.\cite{Zhi2020} Charge neutrality is thereby enforced implicitly, through the quantum charge $n$ of Eq.~(1.6) that closes the self-consistent loop. In the laboratory frame the source electrochemical potential stays pinned at $\mu_{0} = 0$ and the bands move, so that $E_{\mathrm{CNP}} = +U$; this frame is plotted in Figure~\ref{fig:S3}, in which the absolute band edges are $E_C - E_{\mathrm{CNP}} + U$ and $E_V - E_{\mathrm{CNP}} + U$. The equivalent local-frame statement holds the bands fixed (\S1.1); that is the frame of Figure~2(b,d,f), whose dashed curve is $\Delta E = -U$, not the reservoir Fermi level.

\subsection*{6.2 Zero-Bias Conductance}

The Landauer formula\cite{Rahman2003,Lundstrom2017} gives the linear-response conductance at zero drain bias:
\begin{equation}
G = 4\,\frac{e^{2}}{h}\int M(E)\left(-\frac{\partial f}{\partial E}\right)\bigg|_{\Delta E}dE, \tag{6.2}
\end{equation}
where the factor of 4 accounts for spin and valley degeneracies and the thermal broadening function is
\begin{equation}
-\frac{\partial f}{\partial E} = \frac{1}{k_{B}T}\,\frac{\exp[(E - \Delta E)/k_{B}T]}{\left\{1 + \exp[(E - \Delta E)/k_{B}T]\right\}^{2}}. \tag{6.2a}
\end{equation}
Because $M(E)$ is the unshifted local-frame mode density, the broadening is centred on the local-frame level $\Delta E = E_{F,\mathrm{local}} = -U$; the equivalent laboratory-frame form uses $M(E-U)$ centred on the pinned $E_{F} = 0$. A thermal-activation floor $G_{\min} = 4(e^{2}/h)\exp(-E_{\mathrm{gap}}/2k_{B}T)$ is used to prevent unphysical zero conductance when the Fermi level is in the band gap. For the BLG BCFET at 300~K, the zero-bias conductance reaches its minimum $G_{\min} = 8.58\times10^{-4}$~S at $V_{b} = -15.0$~V. That is a conductance, not a conductance per unit width; divided by the $W = 1.082~\mu$m implicit width of the oxide $k_{y}$ grid it becomes $7.93\times10^{2}$~S/m, which is the figure quoted in the Results and discussion section of the main text.

\subsection*{6.3 Self-Consistent Channel Potential and Top-of-the-Barrier Model}

The main text describes how the ToB model\cite{Rahman2003,Lundstrom2017} is used to compute the drain current at finite bias. This model gives a clear picture of ballistic transport through a potential maximum, representing the channel as a single electrostatic node whose potential $U$ is established self-consistently by the balance between the gate-controlled Laplacian potential and the mobile charge in the channel. The source and drain contacts are in local thermal equilibrium at their electrochemical potentials $\mu_{1}$ (source) and $\mu_{2} = \mu_{1} - qV_{d}$ (drain). Each contact is a reservoir in local equilibrium, and at zero drain bias the electrochemical potential is spatially uniform, so the source reservoir and the channel share a single Fermi level~\cite{Rahman2003}. All energies are referred to the laboratory frame of \S1.1, in which the grounded source fixes $\mu_{1} = \mu_{0} = 0$ and the band shift is carried by $U$ inside the Fermi functions of Eq.~(1.6). An applied drain bias lowers only the drain reservoir, giving $\mu_{2} = -qV_{d}$, so the positive-velocity states at the top of the barrier are filled by the source ($\mu_{1}=0$) and the negative-velocity states by the drain ($\mu_{2}=-qV_{d}$)~\cite{Rahman2003}. Substituting the local-frame value $E_{F,\mathrm{local}} = \mu_{0}-U$ for $\mu_{1}$ here would apply the band shift twice and is not done. The channel potential $U$ adjusts to maintain charge neutrality at the top of the barrier. The Laplacian potential $U_{L}$ is the potential at the channel node in the absence of mobile charge; it depends only on the capacitive coupling among the gate, drain, and source:
\begin{equation}
U_{L} = -q\,(\alpha_{G} V_{G} + \alpha_{D} V_{D} + \alpha_{S} V_{S}), \tag{6.3}
\end{equation}
where $\alpha_{G}$, $\alpha_{D}$, and $\alpha_{S}$ are the capacitive coupling ratios to the gate, drain, and source, respectively. The total self-consistent potential is the sum of the Laplacian potential and the energy required to charge the mobile carriers:
\begin{equation}
U = U_{L} + U_{0}\,(N - N_{0}), \tag{6.4}
\end{equation}
where $U_{0} = e^{2}/C_{\Sigma}$ is the single-electron charging energy of the channel capacitor, $C_{\Sigma} = \bar{C}_{t} + \bar{C}_{b}$ is the geometric charging capacitance of Eq.~(1.5), used for both devices (\S1.1), $N_{0} = \int D(E)\,f(E, E_{F})\,dE$ is the equilibrium carrier density at zero drain bias, and $N$ is the non-equilibrium carrier density under finite drain bias.

Figure~S4 illustrates the ToB model used for ballistic transport in the BLG channel. Panel~(a) shows the conduction band edge $E_{c}(z)$ across the source--channel--drain region at equilibrium, identifying the barrier top; the laboratory-frame edge is $E_{c0} + U(z)$, where $E_{c0} = E_{C} - E_{\mathrm{CNP}}$ is the CNP-referenced band edge obtained from $H_{0}(\Delta)$ and $U(z) = -q\psi_{S}(z)$ is the top-of-the-barrier potential energy of Eqs.~(1.7) and~(6.4). Panel~(b) depicts thermionic injection from the source and drain contacts: the probability of emission over the barrier is $\exp(-E_{\mathrm{SB}}/k_{B}T)$ from the source and $\exp(-E_{\mathrm{DB}}/k_{B}T)$ from the drain, so that $I_{\mathrm{LR}} \propto \exp(-E_{\mathrm{SB}}/k_{B}T)$ and $I_{\mathrm{RL}} \propto \exp(-E_{\mathrm{DB}}/k_{B}T)$, where $E_{\mathrm{SB}}$ and $E_{\mathrm{DB}}$ are the barrier heights from the source and from the drain, respectively, to the top of the barrier.\cite{Lundstrom2017} Panel~(c) shows the net drain current $I_{\mathrm{DS}} = I_{\mathrm{LR}} - I_{\mathrm{RL}}$ under applied bias. Panel~(d) illustrates the effect of drain voltage $V_{\mathrm{DS}}$ on the barrier shape; in the absence of drain-induced barrier lowering, $E_{\mathrm{DB}} = E_{\mathrm{SB}} + qV_{\mathrm{DS}}$.

\begin{figure}[H]
    \centering
    \includegraphics[width=0.80\linewidth]{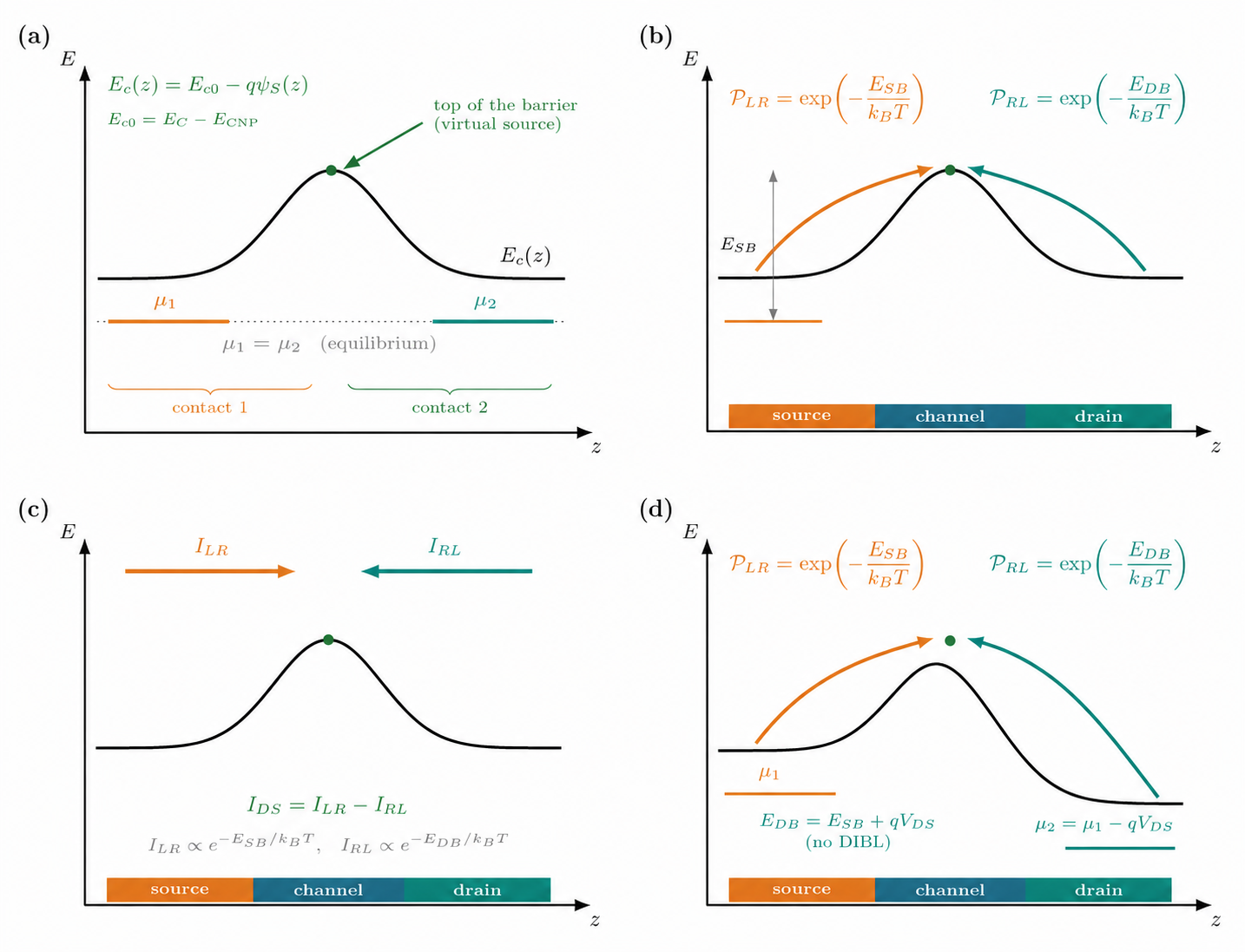}
    \caption{Top-of-the-barrier (ToB) model for ballistic transport in the BLG channel. (a) Energy band diagram showing $E_{c}(z)$ across source--channel--drain at equilibrium. (b) Thermionic injection over the barrier, with emission probabilities $\exp(-E_{\mathrm{SB}}/k_{B}T)$ from the source and $\exp(-E_{\mathrm{DB}}/k_{B}T)$ from the drain. (c) Net drain current $I_{\mathrm{DS}} = I_{\mathrm{LR}} - I_{\mathrm{RL}}$ under bias. (d) Effect of drain voltage $V_{\mathrm{DS}}$ on barrier shape and injection conditions. Throughout, $\psi_{S}$ is the electrostatic potential at the top of the barrier and the corresponding channel potential energy is $U = -q\psi_{S}$.}
    \label{fig:S4}
\end{figure}

The non-equilibrium carrier density is
\begin{equation}
N = \int D(E)\,\frac{f_{S}(E + U, \mu_{1}) + f_{D}(E + U, \mu_{2})}{2}\,dE. \tag{6.5}
\end{equation}
The factor of $1/2$ arises because the channel density is the average of the populations injected from the source and the drain at the top of the barrier. The density of states $D(E)$ entering Eq.~(6.5) is the full self-consistent four-band BLG DOS computed at the converged $\Delta$ for the given gate bias $V_{b}$ (Eq.~5.1). The L-B formula then gives the drain current by summing the net transmission over the Fermi window defined by $\mu_{1} - \mu_{2} = qV_{d}$:
\begin{equation}
I_{\mathrm{DS}} = \frac{4q}{h}\int M(E)\,[f_{S}(E + U, \mu_{1}) - f_{D}(E + U, \mu_{2})]\,dE, \tag{6.6}
\end{equation}
where $M(E)$ is the energy-resolved mode density (Eq.~5.4) and the factor of 4 accounts for spin and valley degeneracy. The shift from $E$ to $E + U$ in the integrand captures how the self-consistent potential rigidly shifts the channel band structure, a key mechanism of the ToB model. To prevent the minimum current from dropping below the thermal-activation floor $I_{\min} = G_{\min} V_{d}$, with $G_{\min} = 4(e^{2}/h)\exp(-E_{\mathrm{gap}}/2k_{B}T)$, thermionic emission over the transport band gap is enforced. This prevents the calculated current from falling below the thermal-activation floor when $\Delta E$ lies deep in the gap.

\section*{7. Subthreshold Swing and Current On/Off Ratio Extraction}

The SS is the voltage required to change the drain current by one decade\cite{Rahman2003}:
\begin{equation}
SS = \left[\frac{d(\log_{10}|I_{d}|)}{dV_{b}}\right]^{-1}\times 10^{3} \quad [\mathrm{mV/dec}]. \tag{7.1}
\end{equation}
A tangent-line method is used to extract $\mathrm{SS_{min}}$. The $\log_{10}|I_{d}|$ vs $V_{b}$ curve is first smoothed with a five-point moving average within a user-defined voltage window on one side of the charge-neutrality point (CNP), defined as the $V_{b}$ value where $|\Delta E - (E_C + E_V - 2E_{\mathrm{CNP}})/2|$ is minimized. The local $SS$ is then computed as the ratio $dV_{b}/d(\log_{10}|I_{d}|)$, and the steepest point (minimum $|SS|$) is found; $\mathrm{SS_{min}}$ is the tangent line through that point. The three samples at each end of the window are discarded before the minimum is taken, because the derivative there is a one-sided difference: the slope it returns is a boundary artifact of the finite window rather than a property of the device, and it is systematically too steep.

The $I_{\mathrm{on}}/I_{\mathrm{off}}$ ratio is measured over a wider voltage window: $\pm1.2$ to $\pm3$~V for the BLG NC-BCFET depending on $L_{b}$ ($\pm1.2$~V at $L_{b} = 8$~nm, $\pm1.8$~V at 10~nm, $\pm2$~V at 12~nm, $\pm2.5$~V at 14~nm and $\pm3$~V for $L_{b} \ge 16$~nm), and $\pm19.5$~V for the BLG BCFET, as the ratio of the highest to lowest absolute drain current. As a reference, the Boltzmann limit is $SS_{\mathrm{ideal}} = \ln(10)\,k_{B}T/q\times 10^{3} \approx 59.5$~mV/dec at 300~K.

\section*{8. Ferroelectric Thickness Optimization and Capacitance Matching}

\subsection*{8.1 Non-Linear Capacitance}

The dependence of $\mathrm{SS_{min}}$ on $L_{b}$---weakly non-monotonic at low drain bias, with a shallow minimum at $L_{b} = 10$~nm for $V_{d} = 10$ and 20~mV, and monotonically increasing for $V_{d} \ge 30$~mV---is set by a competition among three capacitances in the gate stack: $C_{\mathrm{FE}}$, $C_{\mathrm{ox}}$, and $C_{\mathrm{BLG}}$. Only $C_{\mathrm{ox}} = \varepsilon_{0}\varepsilon_{r,t}/L_{t} = 5.29\times10^{-3}~\mathrm{F/m^{2}}$ is linear, i.e.\ independent of the applied bias. The ratio of $C_{\mathrm{FE}}$ to $C_{\mathrm{BLG}}$ determines the effective body factor $m = 1 + C_{s}(1/C_{\mathrm{ox}} + 1/C_{\mathrm{FE}})$, where $C_{s}$ is the semiconductor (BLG) capacitance and $C_{\mathrm{FE}} < 0$ in the NC regime. This is because $C_{\mathrm{FE}}$ and $C_{\mathrm{BLG}}$ are both strong, non-linear functions of the gate voltage. To switch at less than 60~mV/dec, $|1/C_{\mathrm{FE}}|$ must exceed $1/C_{\mathrm{ox}}$, so that the ferroelectric over-screens the depolarization field and makes the surface potential swing more than the gate voltage.\cite{Salahuddin2008,Alam2019}

\subsection*{8.2 Extracted Capacitances and Figures of Merit}

Table~S3 lists the extracted capacitances at the mid-gap operating point ($V_{b} \approx 0$~V, $V_{d} = 10$~mV) for $L_{b} = 8$--20~nm. These were calculated using the Levanyuk formulation\cite{Levanyuk2016} $C_{\mathrm{FE}} = (\varepsilon_{0}\varepsilon_{b} + 1/\kappa)/t_{\mathrm{FE}}$, where $\kappa = 2\alpha + 12\beta P_{f}^{2} + 30\gamma P_{f}^{4}$ is the Landau curvature.

\begin{table}[H]
\centering
\small
\caption{Extracted capacitances and figures of merit at $V_{b} \approx 0$~V, $V_{d} = 10$~mV. $C_{\mathrm{ox}} = 5.29\times10^{-3}~\mathrm{F/m^{2}}$ (constant). ``NC'' is the fraction of the $V_{b}$ sweep with negative Landau curvature ($C_{\mathrm{FE}} < 0$); ``Stab.\ NC'' is the fraction that is additionally stack-stable ($M > 0$, Case~B of \S3.3). $L_{b} = 10$~nm gives the steepest $\mathrm{SS_{min}}$, while $L_{b} = 12$~nm is the best compromise between steepness and stabilized NC coverage.}
\begin{tabular}{cccccccc}
\hline
$L_{b}$ & $\mathrm{SS_{min}}$ & $I_{\mathrm{on}}/I_{\mathrm{off}}$ & $C_{\mathrm{FE}}$ & $|C_{\mathrm{FE}}|/C_{\mathrm{ox}}$ & $C_{\mathrm{BLG,diff}}$ & NC & Stab.\ NC \\
(nm) & (mV/dec) & & ($\mathrm{F/m^{2}}$) & & ($10^{-3}~\mathrm{F/m^{2}}$) & (\%) & (\%) \\
\hline
8 & 15.26 & 117.6 & $-0.136$ & 25.7 & 0.978 & 76 & 62 \\
10 & 15.08 & 121.0 & $-0.109$ & 20.6 & 1.84 & 92 & 77 \\
12 & 17.17 & 120.9 & $-0.0909$ & 17.2 & 2.42 & 99 & 85 \\
14 & 19.31 & 120.7 & $-0.0779$ & 14.7 & 2.83 & 100 & 88 \\
16 & 20.46 & 121.4 & $-0.0682$ & 12.9 & 3.14 & 100 & 90 \\
18 & 22.84 & 120.9 & $-0.0606$ & 11.4 & 3.38 & 100 & 93 \\
20 & 25.24 & 121.4 & $-0.0545$ & 10.3 & 3.57 & 100 & 96 \\
\hline
\end{tabular}
\end{table}

\subsection*{8.3 Thickness-Dependent Simulation Results}

To illustrate the impact of ferroelectric thickness on device performance, Figures~S5 and~S6 show the ferroelectric polarization $P_{f}$ versus the ferroelectric voltage $V_{\mathrm{FE}}$ for $L_{b} = 8$~nm and $L_{b} = 20$~nm, respectively. The $L_{b} = 8$~nm case represents a thinner ferroelectric layer approaching the lower-thickness limit of useful NC stabilization, while the $L_{b} = 20$~nm case represents a thicker ferroelectric layer with greater stabilized-NC coverage. As Table~S3 shows, the stabilized-NC fraction decreases toward the thin end of the simulated range; further reducing $L_{b}$ below 8~nm is therefore expected to shrink, and may ultimately eliminate, the stabilized-NC window. Comparing these two extremes with the $L_{b} = 12$~nm configuration (Table~S3) demonstrates the trade-off between NC amplification strength and stack stability.

\begin{figure}[H]
    \centering
    \includegraphics[width=0.72\linewidth]{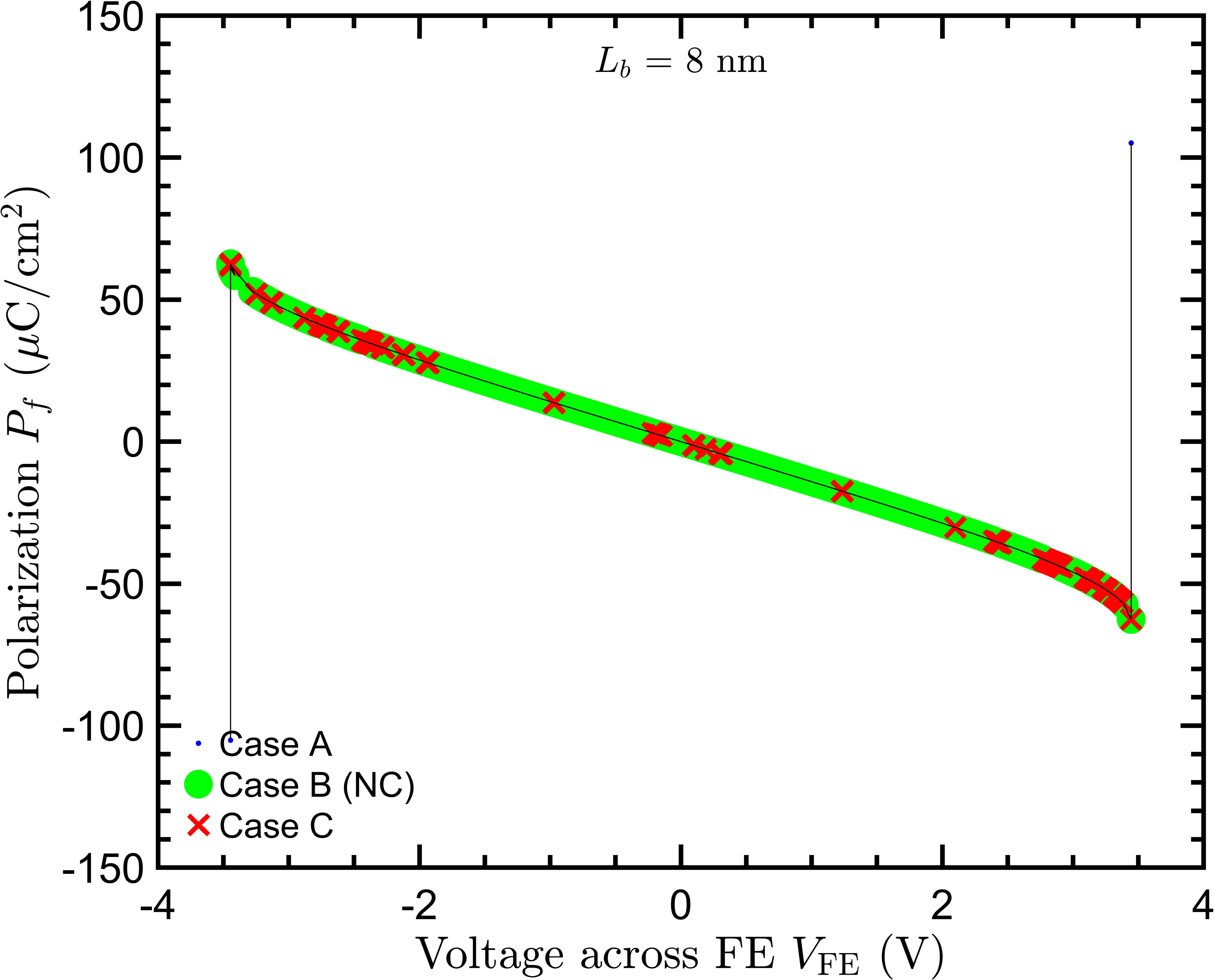}
    \caption{Ferroelectric polarization $P_{f}$ versus ferroelectric voltage $V_{\mathrm{FE}}$ for the BLG NC-BCFET with ferroelectric thickness $L_{b} = 8$~nm, showing the S-shaped $P$--$V$ characteristic and the three operating regimes (Case~A, Case~B [NC], Case~C). The thinner ferroelectric layer yields a smaller fraction of the sweep in the NC regime.}
    \label{fig:S5}
\end{figure}

\begin{figure}[H]
    \centering
    \includegraphics[width=0.72\linewidth]{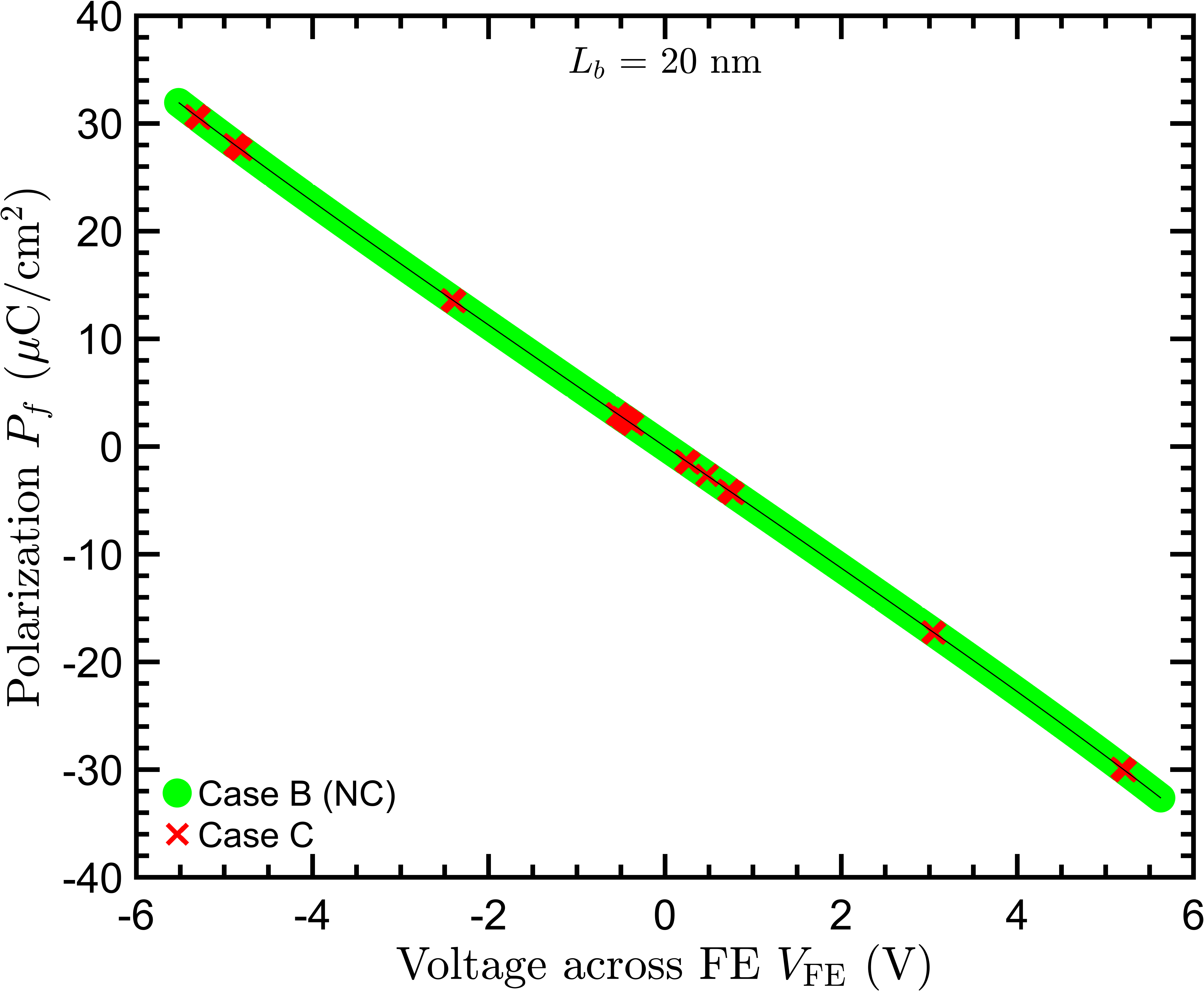}
    \caption{Ferroelectric polarization $P_{f}$ versus ferroelectric voltage $V_{\mathrm{FE}}$ for the BLG NC-BCFET with ferroelectric thickness $L_{b} = 20$~nm, showing the S-shaped $P$--$V$ characteristic with Case~B (NC) and Case~C branches. The thicker ferroelectric layer holds a larger fraction of the sweep in the stabilized NC regime.}
    \label{fig:S6}
\end{figure}

\subsection*{8.4 Optimal Thickness Analysis}

\begin{figure}[H]
    \centering
    \includegraphics[width=0.72\linewidth]{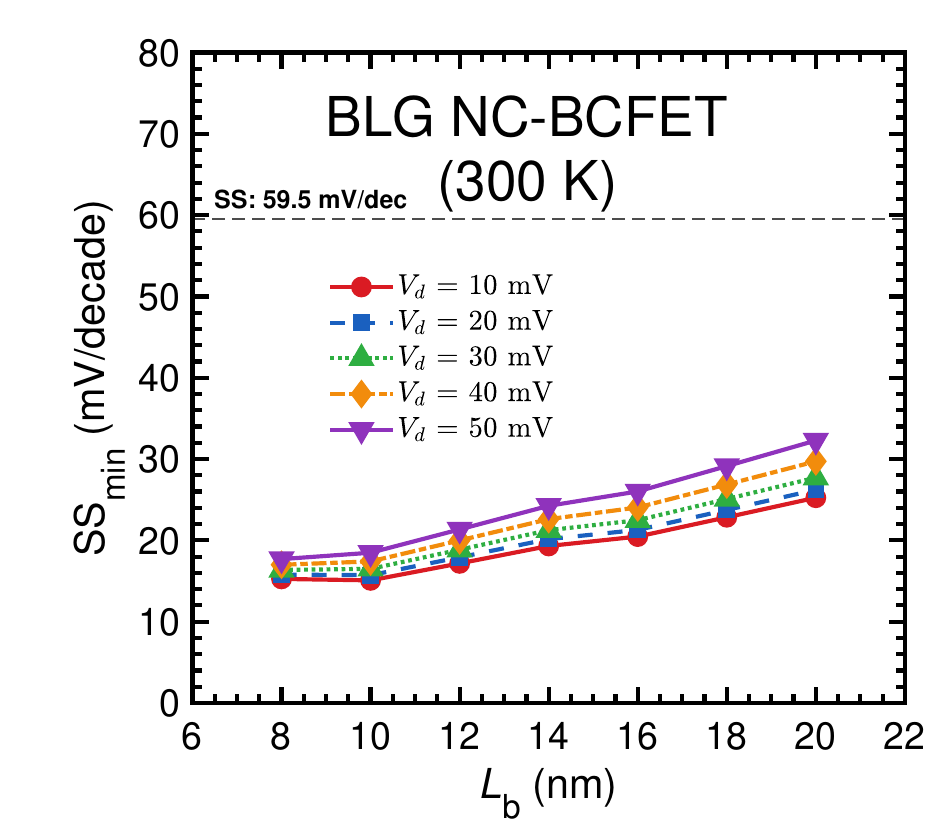}
    \caption{Minimum subthreshold swing $\mathrm{SS_{min}}$ versus ferroelectric thickness $L_b$ for the BLG NC-BCFET at 300~K, at each drain bias $V_d$. The dashed line is the 59.5~mV/dec thermionic limit.}
    \label{fig:S7}
\end{figure}

Figure~\ref{fig:S7} summarizes the drain-bias dependence of the minimum subthreshold swing as a function of ferroelectric thickness and provides the graphical basis for the optimal-thickness analysis below.

In Figure~\ref{fig:S7}, $L_b$ varies from 8 to 20~nm for Al$_{0.55}$Sc$_{0.45}$N; the Landau coefficients follow Gu \textit{et al.}~\cite{Gu2024}, while $\varepsilon_b=3.9$ is an explicit background-permittivity parameter of the adopted model. At $V_d=10$~mV, $\mathrm{SS_{min}}$ values are 15.26, 15.08, 17.17, 19.31, 20.46, 22.84, and 25.24~mV/dec for $L_b=8$, 10, 12, 14, 16, 18, and 20~nm, respectively. The minimum is therefore at 10~nm rather than 8~nm, although the difference is small. The stored negative-curvature/positive-margin coverage increases from 62\% at 8~nm to 96\% at 20~nm.

The ferroelectric is thin and highly non-linear at $L_{b} = 8$~nm: $|C_{\mathrm{FE}}| = 0.136~\mathrm{F/m^{2}}$ ($25.7\times C_{\mathrm{ox}}$). The price is coverage: the Landau curvature is negative over only 76\% of the $V_{b}$ sweep and only 62\% of the sweep satisfies the stack-stability condition $M > 0$, so a substantial part of the sweep leaves the stabilized NC regime. Both coverages improve monotonically with thickness---92\%/77\% at $L_{b} = 10$~nm, 99\%/85\% at $L_{b} = 12$~nm, and 100\%/96\% at $L_{b} = 20$~nm---while the non-linear Landau stiffness weakens, lowering $|C_{\mathrm{FE}}|/C_{\mathrm{ox}}$ (25.7 at 8~nm, 17.2 at 12~nm, 14.7 at 14~nm and 10.3 at 20~nm) and raising $\mathrm{SS_{min}}$ (17.17, 19.31 and 25.24~mV/dec at 12, 14 and 20~nm, respectively). The steepest switching of the series is not at the thinnest ferroelectric: at $V_{d} = 10$~mV the two thinnest devices give $\mathrm{SS_{min}} = 15.26$~mV/dec (8~nm) and 15.08~mV/dec (10~nm), a difference of 0.18~mV/dec, about 1\% of either value. The 10~nm point is therefore best read as the flat bottom of the trend rather than a sharply resolved optimum; the thickness dependence becomes pronounced only from 12~nm upward, where $\mathrm{SS_{min}}$ rises monotonically to 25.24~mV/dec at 20~nm. There is therefore a genuine trade-off rather than a single optimum: the 8--10~nm pair maximizes the NC voltage gain and gives the lowest $SS$ but holds only 62--77\% of the sweep in the stabilized NC regime, whereas $L_{b} = 12$~nm is the thinnest ferroelectric that keeps the Landau curvature negative over essentially the entire sweep (99\%) while still delivering $\mathrm{SS_{min}} = 17.17$~mV/dec, and is thus the best compromise between amplification strength and stabilized single-domain operation. The quantity tabulated as $C_{\mathrm{BLG,diff}}$ is the differential interlayer response of the channel charge, used as a fast criterion for ferroelectric root selection. Because Eqs.~(2.1)--(2.3) carry no channel-potential term---the common mode is held entirely by $U$---the only potential-like variable available there is $\Delta$, and differentiating with respect to it gives $C_{\mathrm{BLG,diff}} = \bar{C}_{t} - \bar{C}_{b} = \varepsilon_{0}\varepsilon_{r,t}/L_{t} - \varepsilon_{0}\varepsilon_{b}/t_{\mathrm{FE}}$: raising $\Delta$ adds charge to one layer and removes it from the other, so the \emph{total} charge responds to the difference of the two gate capacitances. It is therefore a purely geometric quantity, carrying no density-of-states dependence, and its variation with $L_{b}$ is simply the $1/t_{\mathrm{FE}}$ of the bottom dielectric, from $9.78\times10^{-4}~\mathrm{F/m^{2}}$ ($L_{b} = 8$~nm) to $3.57\times10^{-3}~\mathrm{F/m^{2}}$ ($L_{b} = 20$~nm). The quantum capacitance of the channel is never imposed as a lumped element in this work: it enters exactly through the DOS-weighted charge integral of Eq.~(1.6), so that $C_{Q} = -e^{2}\,dn/dU$ appears in the Jacobian of the $U$-solve as $dU/dU_{L} = (1 + C_{Q}/C_{\Sigma})^{-1}$, the standard top-of-the-barrier result.\cite{Datta2005,Rahman2003} Because $\bar{C}_{t} = \bar{C}_{b}$ at $t_{\mathrm{FE}} = \varepsilon_{b}L_{t}/\varepsilon_{r,t} = 6.5$~nm, just below the thinnest ferroelectric considered here, $1/C_{\mathrm{BLG,diff}}$ is large and varies rapidly at the thin end of the sweep; the stability classification in that region should be read qualitatively. The body factor $m$---and thus $SS$---does not follow $L_{b}$ geometrically, because $C_{\mathrm{FE}}$ and $C_{\mathrm{BLG}}$ depend non-linearly on the self-consistent electrostatic solution (through $P_{f}$ and $\Delta$, respectively), while $C_{\mathrm{ox}}$ is a fixed geometric quantity. This thickness trend is qualitatively consistent with the experiments of McGuire \textit{et al.}\cite{McGuire2017}, who showed that a $\sim$12~nm HfZrO$_2$ gate in MoS$_2$ NC-FETs gave the highest internal voltage gain ($\sim$28$\times$) and the steepest $SS$ (6.07~mV/dec). The ferroelectric capacitance magnitude at the optimal thickness in their devices was very close to the series combination of oxide and channel capacitances, maximizing the voltage amplification across the NC layer. The analogy applies to the Al$_{0.55}$Sc$_{0.45}$N/BLG system studied here: the Landau coefficients of Al$_{0.55}$Sc$_{0.45}$N ($\alpha \approx -4.44\times10^{8}~\mathrm{m/F}$) are comparable to those of HfZrO$_2$ ($\alpha \approx -6\times10^{8}~\mathrm{m/F}$), resulting in a similar optimal-thickness window. The design principle that emerges for NC-FETs is that the useful ferroelectric thickness window is bounded from below by the requirement that the subthreshold sweep stay in the stabilized NC regime, and from above by the loss of non-linear Landau stiffness, which diminishes the voltage gain.

\subsection*{8.5 Temperature Dependence}

For $L_b=20$~nm at 100~K, the reduced thermal smearing ($k_{B}T = 8.6$~meV vs 25.9~meV at 300~K) sharpens the Fermi--Dirac transition and, combined with NC amplification, gives $\mathrm{SS_{min}} = 2.50$~mV/dec and $I_{\mathrm{on}}/I_{\mathrm{off}} = 3.92\times10^{5}$ at $V_{d} = 10$~mV, falling to $1.31\times10^{5}$ at $V_{d} = 50$~mV. The BLG NC-BCFET is unique in that it can exploit both NC voltage gain and cryogenic thermal sharpening simultaneously. In conventional NC-FETs with fixed band-gap semiconductors, lowering the temperature sharpens the Fermi window but does not change the channel band structure. In BLG, by contrast, the gate-tunable gap itself responds more steeply to the amplified surface potential at lower temperatures.

\section*{9. Simulation Parameters}

The real roots of the Landau polynomial, Eq.~(3.4), are obtained as the eigenvalues of its companion matrix.\cite{Quarteroni2014}

\begin{table}[H]
\centering
\footnotesize
\caption{Simulation parameters for the BLG BCFET and BLG NC-BCFET.}
\begin{tabular}{lcc}
\hline
Parameter & BLG BCFET& BLG NC-BCFET\\
\hline
$V_{b}$ sweep range & $-20$ to $+20$~V & $-5$ to $+5$~V \\
Number of $V_{b}$ points & 16{,}000 & 8{,}000 \\
$V_{t}$ (fixed) & $+5$~V & $+5$~V \\
Bottom dielectric & SiO$_2$ ($\varepsilon_r = 3.9$, $L_b = 20$~nm) & Al$_{0.55}$Sc$_{0.45}$N ($\varepsilon_b = 3.9$, $t_{\mathrm{FE}} = 8$--20~nm) \\
Top dielectric & HfO$_2$ ($\varepsilon_r = 5.98$, $L_t = 10$~nm) & HfO$_2$ ($\varepsilon_r = 5.98$, $L_t = 10$~nm) \\
$k$-grid $N_k \times N_k$ & $500 \times 500$ & $1000 \times 1000$ \\
Energy grid $N_E$ & 1{,}000 & 500 \\
Gaussian broadening $\gamma$ & 1.5~meV & 1.5~meV \\
Self-consistent tolerance & $10^{-6}$~eV & $10^{-6}$~eV \\
Max outer SCF iterations & 100 & 100 \\
Drain voltages $V_d$ & 10, 20, 30, 40, 50~mV & 10, 20, 30, 40, 50~mV \\
Temperature & 300~K & 300~K and 100~K \\
Floating-point precision & 64-bit & 64-bit \\
\hline
\end{tabular}
\end{table}

\section*{10. Physical Constants and Material Parameters}

\begin{table}[H]
\centering
\small
\caption{Physical constants and material parameters used in the simulations.\cite{Gu2024,McCann2013}}
\begin{tabular}{lcl}
\hline
Symbol & Value & Description \\
\hline
$e$ & $1.602\times10^{-19}$~C & Elementary charge \\
$h$ & $6.626\times10^{-34}$~J\,s & Planck constant \\
$\hbar$ & $1.055\times10^{-34}$~J\,s & Reduced Planck constant \\
$k_{B}$ & $1.381\times10^{-23}$~J/K ($8.617\times10^{-5}$~eV/K) & Boltzmann constant \\
$\varepsilon_{0}$ & $8.854\times10^{-12}$~F/m & Vacuum permittivity \\
$a_{cc}$ & 1.42~\AA & Carbon--carbon bond length \\
$a$ & $\sqrt{3}\,a_{cc} = 2.46$~\AA & Lattice constant \\
$c_{0}$ & 3.35~\AA & Interlayer spacing \\
$\gamma_{0}$ & 3.16~eV & Nearest-neighbor hopping \\
$\gamma_{1}$ & 0.39~eV & Interlayer coupling (A$_2$--B$_1$ dimer) \\
$\gamma_{3}$ & 0.38~eV & Trigonal warping \\
$\gamma_{4}$ & 0.14~eV & Electron--hole asymmetry \\
$v_{F}$ & $1.02\times10^{6}$~m/s & Fermi velocity \\
$\alpha_{\mathrm{FE}}$ & $-4.44\times10^{8}$~m/F & Landau 1st-order (negative $\rightarrow$ NC) \\
$\beta_{\mathrm{FE}}$ & $3.73\times10^{7}~\mathrm{m^{5}\,F^{-1}\,C^{-2}}$ & Landau 2nd-order \\
$\gamma_{\mathrm{FE}}$ & $1.55\times10^{8}~\mathrm{m^{9}\,F^{-1}\,C^{-4}}$ & Landau 3rd-order \\
$x$ (Sc) & 0.45 & Scandium concentration \\
$x_{c}$ & 0.6115 & Critical Sc concentration \\
\hline
\end{tabular}
\end{table}

\bibliography{references}